%% file: main.tex
\documentclass{article}
\usepackage[a4paper, margin=1.3in]{geometry}
\RequirePackage{amsthm,amsmath,amsfonts,amssymb,commath,bbm}
\usepackage{amsmath} 
\usepackage{hyperref}
\usepackage{cleveref}

\usepackage{graphicx}
\usepackage{subcaption}

\usepackage{todonotes}

\usepackage{fancyhdr}
\usepackage{algpseudocode}% algorithmicx
\usepackage{algorithm}% 

\usepackage{enumerate}

\newtheorem{theorem}{Theorem}[section]
\newtheorem{lemma}{Lemma}[section]
\newtheorem{corollary}{Corollary}[section]
\newtheorem{proposition}{Proposition}[section]

\newtheorem{definition}{Definition}[section]
\newtheorem{assumption}{Assumption}[section]

\Crefname{assumption}{Assumption}{Assumptions}
\newtheorem{example}{Example}[section]
\Crefname{example}{Example}{Examples}
\newtheorem{remark}{Remark}[section]

\usepackage[round,authoryear]{natbib}
\newcommand{\gap}{\mathrm{gap}}
\newcommand{\osc}{\mathrm{osc}}
\newcommand{\ELL}{\mathrm{L}}

\newcommand{\KF}{K_{\mathrm F}^*}
\newcommand{\LKF}{\widehat{K_{\mathrm F}^*}}
\newcommand{\OKF}{K_{\mathrm F}^\star}
\newcommand{\OLKF}{\widehat{K_{\mathrm F}^\star}}

\newcommand{\LKIF}{\widehat{K_{\mathrm{IF}}^*}}
\newcommand{\OKIF}{K_{\mathrm{IF}}^\star}
\newcommand{\OLKIF}{\widehat{K_{\mathrm{IF}}^\star}}

\newcommand{\BF}{\beta_{\mathrm F}}

\newcommand{\BIF}{\beta_{\mathrm{IF}}}
\newcommand{\LBIF}{\widehat{\beta_{\mathrm{IF}}}}
\newcommand{\OLBIF}{\widehat{\beta^\star_{\mathrm{IF}}}}

\newcommand{\LBI}{\widehat{\beta_{\mathrm{I}}}}
\newcommand{\OLBI}{\widehat{\beta^\star_{\mathrm{I}}}}

\fancypagestyle{firstpage}{%
  \lhead{}
  \rhead{}
}

\title{\vspace{-1cm} \textbf{Limitations of nonreversible lifts for subgeometric diffusions}\\
        }

\author{\href{mailto:Mengxi.Gao@warwick.ac.uk}{Mengxi Gao }\\University of Warwick
\and \href{mailto:gareth.o.roberts@warwick.ac.uk}{Gareth O. Roberts}\\University of Warwick
\and \href{mailto:andi.wang@warwick.ac.uk}{Andi Q. Wang} \\University of Warwick}
\date{\vspace{-5ex}} %NO DATE

\begin{document}
\maketitle
\thispagestyle{firstpage}
\pagestyle{firstpage}

\setlength{\parskip}{0.7em}
\setlength{\parindent}{0pt}

\begin{abstract}
    We study the effect of lifting on continuous-time Markov processes with potentially subgeometric convergence rates. Using a series of weak Poincar\'e-type inequalities, we derive a bound on the maximum acceleration that can be obtained through lifting. We present several examples to explore the sharpness of our derived bound, where 
    the resulting bound has the same square-root order as in the spectral gap setting.
    Our results provide a framework for understanding the potential and limitations of non-reversible lifting beyond geometric convergence.
\end{abstract}

\input{261113}

% \bibliographystyle{IEEEtran}
% \bibliography{IEEEabrv,references}

\bibliography{ref}

\end{document}

%% file: 261113.tex
\section{Introduction}
Nonreversible Markov processes have attracted considerable interest as a means to potentially accelerate the convergence of reversible Markov processes to stationarity, by reducing the diffusive behaviour commonly observed among reversible processes.  The diffusive-to-ballistic speed-ups for nonreversible Markov processes have been investigated in various models, including the underdamped Langevin dynamics \cite{lelievre2013optimal}, the Zig-Zag process \cite{bierkens2019zig}, and the Bouncy Particle Sampler \cite{bouchard2018bouncy}. 

Inspired by the notion of lifts of discrete-time Markov chains in \cite{chen1999lifting} and \cite{diaconis2000analysis}, the work of \cite{eberle2024non} builds a second-order lifting structure to show that various well-known nonreversible continuous-time Markov processes used in applications are lifts in this sense of simple reversible diffusions. Furthermore, they introduce the notion of non-asymptotic relaxation times and show that these can be reduced by at most a square root through lifting, thereby establishing fundamental limitations on the speed-up achievable by nonreversible Markov processes through lifting.
Their analysis is restricted to situations when the underlying reversible process possesses a nonzero spectral gap.

In this note, we consider the lifting framework in the setting where the underlying reversible process may not possess a spectral gap. Our goal is to quantify how much acceleration can still be achieved by introducing nonreversibility in this subgeometric regime. Our result thus extends the foundational work of \cite{eberle2024non} beyond exponential decay.

{The main contributions of this work are as follows. First, we introduce \textit{weak iterated Poincar\'e inequalities} \eqref{eq: IWPI}, generalizing the singular value gap used in \cite{eberle2024non}. Second, by investigating optimal weak Poincar\'e inequalities, we 
establish several comparison theorems among weak Poincar\'e-type inequalities (\Cref{Sec: comparison}). Finally, based on these comparisons, we establish in \Cref{thm: main} the maximal acceleration attainable through second-order lifting.}

The rest of this paper is organized as follows. In \Cref{Sec: Lift}, we describe the lifting framework of \cite{eberle2024non}. In \Cref{Sec:WPIs} and \Cref{Sec: comparison} we review the definition and basic properties of weak Poincar\'e inequalities and show how comparison techniques can be used to derive lower bounds within this framework. In \Cref{Sec: Main result} we present our main results together with their proofs. Finally, in \Cref{Sec: Examples}, we discuss several examples to explore the sharpness of the bound that we derived.

\section{Second-order Lift Structure}
\label{Sec: Lift}
We consider a reversible diffusion process with state space $S$ and a  time-homogenous Markov process with state space $\hat S = S \times V$ and invariant probability measures $\mu(\dif x)$ and $\hat \mu (\dif x,\dif v) = \mu(\dif x)\kappa (x, \dif v)$, respectively. We frequently think of the components of the latter as a position together with a velocity or momentum variable. The associated Markov semigroups acting on $\ELL^2(\mu)$ and $\ELL^2(\hat \mu)$ are $(P_t)_{t\geq0}$ and $(\hat P_t)_{t\geq0}$, and their infinitesimal generators are denoted by $(\mathcal L, \mathcal D(\mathcal L))$ and $(\mathcal{\hat L}, \mathcal D(\mathcal{\hat L}))$, respectively. Let $\pi: \hat S\to S , (x, v) \mapsto x$, denote the projection onto the first coordinate.

\begin{definition}\cite[Definition 1]{eberle2024non}
The semigroup $(\hat{P}_t)_{t \geq 0}$ is a second-order lift of $\left(P_t\right)_{t \geq 0}$ if
\begin{align*}
    \mathcal D(\mathcal L) \subseteq\left\{f \in \ELL^2(\mu): f \circ \pi \in \mathcal D(\mathcal{\hat L})\right\},
\end{align*}
and for all $f, g \in \mathcal D(\mathcal L)$ we have
\begin{align*}
    \langle g \circ \pi, -\mathcal{\hat L}(f \circ \pi)\rangle_{\ELL^2(\hat \mu)} &=0,\\
    \frac{1}{2} \langle \mathcal{\hat L}(f \circ \pi), \mathcal{\hat L}(g \circ \pi) \rangle_{\ELL^2(\hat \mu)} &=\langle f, -\mathcal L g\rangle_{\ELL^2(\mu)}.
\end{align*}
\end{definition}

\begin{proposition}\cite[Example 3]{eberle2024non} Given a potential $U \in C^1(\mathbb{R}^d)$ such that $U(x) \rightarrow \infty$ as $|x|\rightarrow \infty$ and  $\int \exp(-U (x))\dif x < \infty$, consider the following three process:
\begin{itemize}
    \item the \emph{overdamped Langevin diffusion}, with state space $S = \mathbb{R}^d$, given by the SDE  
\begin{align*}
    \dif Z_t = -\frac{1}{2}\nabla U (Z_t ) \dif t + \dif B_t,
\end{align*} 
with invariant measure $\mu(\dif x) \propto  \exp(-U (x)) \dif x$, where $(B_t)_{t\geq 0}$ is a $d$-dimensional standard Brownian motion.
\item \emph{Underdamped Langevin dynamics}: the diffusion process $(X_t,V_t)$ on $\hat{S} = \mathbb{R}^d \times \mathbb R^d$ solving the SDE  
\begin{align*}
    \dif X_t & =V_t \dif t \\
\dif V_t & =-\nabla U\left(X_t\right) \dif t-\gamma V_t \dif t+\sqrt{2 \gamma} \dif B_t,
\end{align*}
with invariant measure is $\hat \mu = \mu \otimes \kappa$,
with friction parameter $\gamma\ge 0$.
\item The \emph{Zig-Zag process} $(X_t, V_t)$ is a piecewise-deterministic Markov process with state space $\hat{S} = \mathbb{R}^d \times \{-1,+1\}^d$, and the generator of the associated transition semigroup in $\ELL^2(\hat{\mu})$ is given by
\begin{align*}
\hat{\mathcal{L}} f(x, v)=v & \cdot \nabla_x f(x, v)+\sum_{i=1}^d\left(v_i \partial_i U(x)\right)_{+}\left(f\left(x, v-2 e_i v_i\right)-f(x, v)\right) \\
& +\gamma\left(\int f(x, w) \kappa(\dif w)-f(x, v)\right),
\end{align*}
for compactly supported functions $f \in C^1(\mathbb{R}^d)$ and $(x, v) \in \mathbb{R}^d \times \{-1,+1\}^d$.
\end{itemize}

Then, the underdamped Langevin dynamics and  the Zig-Zag process are all second-order lifts of the overdamped Langevin process.
\end{proposition}

\section{Weak Poincaré inequalities}
\label{Sec:WPIs}

Let $(\mathsf{E}, \mathcal{F}, \mu)$ be a probability space and $(\mathcal L, \mathcal D(\mathcal L))$ a densely-defined linear operator which generates a Markov semigroup $(P_t)_{t\geq 0}$ on $\ELL^2(\mu)$, where $\mu$ is its invariant measure, $\mathrm{L}^2(\mu)$ is the Hilbert space of the set of real-valued measurable functions that are square integrable with respect to $\mu$, equipped with the inner product $\langle f, g\rangle_\mu=\int_{\mathrm{E}} f(x) g(x) \mathrm{d} \mu(x)$, for $f, g \in \mathrm{L}^2(\mu)$, let $\|f\|^2_{\ELL^2(\mu)} = \langle f, f\rangle_{\ELL^2(\mu)}$, and  write $\mathrm{L}^2_0(\mu)$ as the set of functions $f \in \mathrm{L}^2(\mu)$ satisfying $\int f \dif \mu= 0$. 

The standard Poincar\'e inequality (SPI) for $\mathcal L$ is
\begin{align}
\label{eq: SPI}
    C_{\mathrm{P}} \left\| f\right\|_{\ELL^2(\mu)}^2  &\leq \left\langle f,-{\mathcal{L}} f\right\rangle_{\ELL^2(\mu)},
\end{align}
for all $f\in \ELL^2_0(\mu) \cap \mathcal D (\mathcal L)$. 
It is well known that \Cref{eq: SPI} is equivalent to the exponential $\ELL^2$-convergence of $P_t$: for all $f\in \ELL^2_0(\mu)$ and $t\geq 0$, 
\begin{align*}
    \left\| P_tf\right\|_{\ELL^2(\mu)}^2 \leq \exp(-2tC_{\mathrm{P}}) \|f\|_{\ELL^2(\mu)}^2.
\end{align*}

We define the spectral gap of $\mathcal L$ to be 
\begin{align*}
    \gap(\mathcal L) := \inf_{\alpha \in \sigma\left(-\mathcal L|_{\ELL^2_0(\mu)}\right )} \mathrm{Re}(\alpha),
\end{align*} 
$\sigma$ denoting the spectrum of an operator.

When $\mathcal L$ is self-adjoint, we have a variational characterization of the spectral gap,
\begin{align}
    \gap(\mathcal L)=\inf_{f \in \mathcal D(\mathcal L) \cap \ELL^2_0(\mu)}\frac{\left\langle f,-{\mathcal{L}} f\right\rangle_{\ELL^2(\mu)}}{\|f\|^2_{\ELL^2(\mu)}}.
\end{align}

For the transition semigroup $(P_t)_{t\geq0}$ of a potentially nonreversible Markov process with invariant probability measure $\mu$ and generator $(\mathcal{L}, \mathcal D(\mathcal{L}))$ in $\ELL^2(\mu)$, the process may not satisfy a Poincar\'e inequality, yet still converge exponentially fast. \cite{eberle2025convergence} show that a \textit{flow Poincar\'e inequality} is equivalent to exponential decay of the $\ELL^2(\mu)$-norm averaged over time $T$ along trajectories of the semigroup.

\begin{definition}\cite[Definition 8]{eberle2025convergence}
    The semigroup $(P_t)_{t\geq 0}$ satisfies a \emph{flow Poincar\'e inequality (FPI)} with constant $\nu^{-1}\in(0,\infty)$  and duration $T\in(0,\infty)$ if  
    \begin{align*}
        \int_0^T\left\| P_tf\right\|_{\ELL^2(\mu)}^2 \dif t &\leq \frac{1}{\nu} \int_0^T\left\langle P_tf,-\mathcal{L}P_tf\right\rangle_{\ELL^2(\mu)} \dif t
    \end{align*}
    for all $f \in \ELL^2_0(\mu) \cap \mathcal D(\mathcal L)$.
\end{definition}
This can be seen as a Poincar\'e inequality integrated over a fixed time length $T \in (0, \infty)$ along trajectories of the semigroup.

\begin{proposition}\cite[Theorem 9]{eberle2025convergence}
    Suppose the semigroup $(P_t)_{t\geq 0}$ satisfies a FPI with constant $\nu^{-1}\in(0,\infty)$  and duration $T\in(0,\infty)$. Then for all $f \in \ELL^2_0(\mu)$ and $t \geq 0$, one has  
    \begin{align*}
        \int_t^{T+t} \|P_sf\|_{\ELL^2(\mu)}^2 \dif s \leq e^{-2\nu t}\int_0^T \|P_sf\|_{\ELL^2(\mu)}^2 \dif s,
    \end{align*}
    which implies $\|P_tf\|_{\ELL^2(\mu)} \leq Ce^{-\nu t}\|f\|_{\ELL^2(\mu)}$, with $C=e^{\nu T}$.
\end{proposition} 

Our main interest lies in the case where the generator $\mathcal{L}$ does not admit a spectral gap. In this case, the flow Poincar\'e inequality cannot hold. To tackle the absence of spectral gap, we employ the framework of \textit{weak Poincar\'e inequalities} in the context of continuous-time Markov processes as developed in \cite{rockner2001weak}; see \cite{andrieu2022comparison} for analogous discrete-time developments. 
 
\begin{definition}
\cite[Weak Poincar\'e inequality]{rockner2001weak, andrieu2022comparison}
\label{def:WPI}
    The semigroup $(P_t)_{t\geq 0}$ satisfies a  {\emph{$\beta$-weak Poincar\'e inequality (WPI)}} if for any $f \in \ELL^2_0(\mu)\cap \mathcal D(\mathcal{L})$, 
    \begin{align}
        \left\| P_tf\right\|_{\ELL^2(\mu)}^2  &\leq s \left\langle P_tf,-{\mathcal{L}} P_tf\right\rangle_{\ELL^2(\mu)}  + \beta(s)\|f\|^2_\osc,\quad\forall s>0,
        \label{eq:WPI_beta}
    \end{align}
    where $\|f\|^2_\osc$ is the oscillation seminorm defined by $\|f\|_\osc:=\mathrm{ess}_\mu\sup f-\mathrm{ess}_\mu\inf f$, where $\mathrm{ess}_\mu\sup f := \inf\{a \in \mathbb{R}: \mu(f^{-1}(a,\infty))=0\}$, $\mathrm{ess}_\mu\inf f := \sup\{b \in \mathbb{R}: \mu(f^{-1}(-\infty,b)) =0\}$, and
    {$\beta\in \mathcal{B}$, 
    with $\mathcal{B}$ denoting the set of admissible $\beta$ functions defined by $$\mathcal{B}:=\{\beta :(0,\infty)\to [0,\infty)\mid \beta \text{ is a decreasing function with} \lim_{s \rightarrow \infty} \beta(s) = 0\}.$$ }  
 \end{definition}

    \begin{remark}
        For any Markov semigroup $(P_t)_{t \geq 0}$ on $\ELL^2(\mu)$, $f \in \ELL^2(\mu)$, we have $\|P_tf\|_\osc^2\leq \|f\|_\osc^2$. Moreover, without loss of generality, $\beta$ can be chosen such that $\beta(s) \leq 1/4$ for all $s > 0$, since $\|f\|^2_{\ELL^2(\mu)}\leq \|f\|^2_\osc/4$ for any $f \in \ELL^2_0(\mu)$ by Popoviciu's inequality.
    \end{remark}
    
        We also note that most formulations of the WPI require \Cref{eq:WPI_beta} being satisfied for all test functions $g \in \ELL_0^2(\mu)\cap \mathcal D(\mathcal L)$, whereas we have restricted to functions $g=P_tf$ as this will suffice for our purposes.

    \begin{definition}
    \label{thm:K*}
        For $\beta$ as above, let
    $K : [0, \infty) \rightarrow [0, \infty)$ be given by $ K(u):= \mathbbm{1}_{\{u>0\}}u \cdot \beta(1 / u)$.
    Then let $K^* : [0, \infty) \rightarrow [0, \infty]$ be defined as the convex conjugate,
    \begin{align*}
        K^*(v) := \sup_{u\geq 0}\{uv - K(u)\}.
    \end{align*} 
    {Let $\mathcal{C}$ be the set of the admissible $K^*$ functions defined as
    \begin{align}
        \mathcal{C}:=\left \{K^*:[0,\infty) \to [0,\infty]\mid  {K}^*(v)=\sup_{u\geq 0}\{uv - u\beta(1/u)\}, \beta \in \mathcal{B}\right \}.
        \label{eq:C_def}
    \end{align}}
    Then for $f\in \ELL^2_0(\mu)$ such that $0<\|f\|^2_\osc<\infty$, the weak Poincar\'e inequality \eqref{eq:WPI_beta} can be formulated as follows (with $u=1/s>0$):
    \begin{align*}
        u\|P_tf\|^2_{\ELL^2(\mu)}\leq \left\langle P_tf,-{\mathcal{L}}P_tf\right\rangle_{\ELL^2(\mu)}+K(u)\|f\|^2_\osc,
    \end{align*}
     which by rearranging terms can be expressed as an \emph{optimized $K^*$-WPI}:
\begin{align}
    K^*\left(\frac{\left\| P_tf\right\|_{\ELL^2(\mu)}^2 }{\|f\|^2_\osc}\right)&\leq \frac{\left\langle P_tf,-{\mathcal{L}}P_tf\right\rangle_{\ELL^2(\mu)} }{\|f\|^2_\osc}.
\end{align}

\end{definition}

\begin{theorem}
    \cite[Theorem 8]{andrieu2022comparison}
\label{thm:WPI_convergence}
    Suppose a WPI with $\beta: (0, \infty) \rightarrow [0, \infty)$ holds for $(P_t)_{t\geq 0}$. Then, for any $f \in \ELL^2_0(\mu)\cap \mathcal D(\mathcal L)$ such that $\|f\|^2_\osc < \infty$ and any $t>0$,
    \begin{align}
    \label{eq:convergence rate}
    \|P_tf \|^2_{\ELL^2(\mu)} \leq \|f\|^2_\osc \gamma (t),
    \end{align}
    where $F: \left(0, 1/4\right] \rightarrow \mathbb{R}$ is the decreasing convex and invertible function $F(x) :=\int_x^{1/4} \frac{\dif v}{2K^*(v)}$, with $K^* : [0, \infty) \rightarrow [0, \infty]$ defined as \Cref{thm:K*}, and $\gamma := F^{-1}$ is the inverse function.
\end{theorem}

The proof for the continuous-time processes is essentially similar to the proof of \cite[Theorem 8]{andrieu2022comparison} and is given in \Cref{prf:thm:WPI_convergence} for completeness.

\begin{remark}
\label{rmk:strongtoweak}
    If the generator $\mathcal L$ of a semigroup $(P_t)$ satisfies a standard Poincar\'e inequality with constant $C_{\mathrm{P}}>0$, one may take the corresponding $\beta$ to be $\beta(s)=\mathbbm{1}_{\{s\leq C_{\mathrm{P}}^{-1}\}}$. Conversely, if a WPI with function $\beta(s)=\mathbbm{1}_{\{s\leq C_{\mathrm{P}}^{-1}\}}$ holds, then one can deduce that a standard Poincar\'e inequality  holds.
\end{remark}

\begin{remark}
    One subtlety regarding these functional inequalities is the use of the Dirichlet form $\left\langle f,-{\mathcal{L}} f\right\rangle_{\ELL^2(\mu)}$. As $\mathcal{L}=\mathcal{L}_s+\mathcal{L}_a$, where $\mathcal{L}_s$ and $\mathcal{L}_a$ denote the symmetric and the anti-symmetic part, respectively, we have $\left\langle f,-{\mathcal{L}} f\right\rangle_{\ELL^2(\mu)}=\left\langle f,-{\mathcal{L}_s} f\right\rangle_{\ELL^2(\mu)}$. So for non-constant $f \in \mathrm{Ker}(\mathcal{L}_s)$, inequalities such as the SPI or WPI would fail, while the process may still converge asymptotically to its invariant measure. This issue is addressed by studying \emph{hypocoercivity}, see \cite{villani2006hypocoercivity, dolbeault2009hypocoercivity},
    where convergence arises from the interaction between the symmetric and antisymmetric parts of the generator, rather than from the symmetric part alone. Hence, motivated by the time-averaging in the FPI, it is natural to consider a weakened flow Poincar\'e inequality that can capture convergence behavior beyond the exponential case.
\end{remark}

Following \cite{brigati2024explicit}, we define the \textit{weak flow Poincar\'e inequality} for $(P_t)_{t\geq 0}$ as follows.

\begin{definition}\cite[Weak Flow Poincar\'e inequality]{brigati2024explicit}
    The semigroup $(P_t)_{t\geq 0}$ satisfies a \emph{weak flow Poincar\'e inequality (WFPI)} with $\BF: (0, \infty) \rightarrow [0, \infty)$ and duration $T \in (0, \infty)$ if for any $f \in \ELL^2_0(\mu)\cap \mathcal D(\mathcal L)$ and $s>0$,
    \begin{align*}
        \int_0^T\left\| P_tf\right\|_{\ELL^2(\mu)}^2 \dif t &\leq s \int_0^T\left\langle P_tf,-\mathcal{L}P_tf\right\rangle_{\ELL^2(\mu)} \dif t + \BF(s)\|f\|^2_\osc,
    \end{align*}
    where $\BF$ is as in \Cref{def:WPI}.
    
    By rearranging terms, we can also express the WFPI as an \emph{optimized $\KF$-WFPI}:  
    \begin{align}
    \label{eq:KF}
        \KF\left(\frac{\int_0^T\left\| P_tf\right\|_{\ELL^2(\mu)}^2 \dif t}{\|f\|^2_\osc}\right) &\leq \frac{\int_0^T\left\langle P_tf,-\mathcal{L}P_tf\right\rangle_{\ELL^2(\mu)} \dif t}{\|f\|^2_\osc},
\end{align}
where $\KF=\sup_{u\geq 0}\{uv - u\BF(1/u)\}$.
\end{definition}
\begin{remark}
     Without loss of generality, $\BF(s)$ can be chosen such that $\BF(s) \leq T/4$ for all $s > 0$, since $\|P_tf\|_\osc^2\leq \|f\|_\osc^2$ and $\|f\|^2_{\ELL^2(\mu)}\leq \|f\|^2_\osc/4$ for any $f \in \ELL^2_0(\mu)$.
\end{remark}
\begin{theorem}\cite[Theorem 5.]{brigati2024explicit}
\label{thm:WFPI_convergence}
    Suppose a WFPI with $\BF: (0, \infty) \rightarrow [0, \infty)$ and duration $T \in (0, \infty)$ holds for $(P_t)_{t\geq 0}$. Then, for any $f \in \ELL^2_0(\mu)\cap \mathcal D(\mathcal L)$ such that $\|f\|^2_\osc < \infty$ and any $t>0$,
    \begin{align*}
        \int_t^{T+t} \|P_sf\|_{\ELL^2(\mu)}^2 \dif s \leq F^{-1}(t)\|f\|^2_\osc,
    \end{align*}  where $F: \left(0, T/4\right] \rightarrow \mathbb{R}$ is the decreasing convex and invertible function $F(x) :=\int_x^{T/4} \frac{\dif v}{2K^*(v)}$, with $\KF : [0, \infty) \rightarrow [0, \infty]$ defined as \Cref{eq:KF}.
    
    It follows that $\|P_tf\|_{\ELL^2(\mu)}^2 \leq \frac{1}{T}F^{-1}(t-T)\|f\|^2_\osc$ for any $t>T$.
\end{theorem}

\Cref{thm: from rate to WFPI} shows that we can derive a converse result which recovers a FPI from the contraction.
\begin{proposition}
\label{thm: from rate to WFPI}
Suppose that for all $f \in \ELL_0^2(\mu)$ and $t \geq 0$, one has
\begin{align*}
        \frac{    \int_t^{T+t}\left\| P_{s}f\right\|_{\ELL^2(\mu)}^2 \dif s}{\|f\|^2_\osc} \leq F^{-1}\left(t+F\left(\frac{    \int_0^{T}\left\| P_{s}f\right\|_{\ELL^2(\mu)}^2 \dif s}{\|f\|^2_\osc}\right)\right),
\end{align*}
for a function $F: (0,T/4]\rightarrow [0, \infty)$ which is decreasing, continuous, divergent at 0, with an inverse function $F^{-1}$ which is decreasing, continuous, and convex, and such that $-(\mathrm{D}F)^{-1}$ is convex on $(0,T/4]$.

Then the semigroup $(P_t)_{t\geq0}$ satisfies a weak flow Poincar\'e inequality with $\KF=-\frac{1}{2 }(\mathrm{D}F)^{-1}(v)$ and duration $T$, from which a corresponding $\BF$ can be deduced via convex duality.
\end{proposition}

\cite{brigati2024explicit} only considered this WFPI for undamped Langevin dynamics, while we define it for a general semigroup $(P_t)_{t\geq 0}$. The proofs of \Cref{thm:WFPI_convergence} and \Cref{thm: from rate to WFPI} are provided in \Cref{prf:thm:WFPI_convergence}.

Next, inspired by \cite{eberle2024non, huang2026bernstein}, we define the weak \textit{iterated} Poincar\'e inequality and the weak flow \textit{iterated} Poincar\'e inequality as follows.

\begin{definition}\cite[]{huang2026bernstein}
    The semigroup $(P_t)_{t\geq 0}$ satisfies an \emph{iterated Poincar\'e inequality (IPI)} with $\eta^{-2}$ if for all $f \in \ELL^2_0(\mu) \cap \mathcal D(\mathcal L)$ and $t\ge 0$,
    \begin{align*}
    \left\| P_tf\right\|_{\ELL^2(\mu)}^2  \leq \frac{1}{\eta^2} \|\mathcal{L}P_tf\|^2_{\ELL^2(\mu)}.
    \end{align*}
    Moreover, we define the \emph{singular value gap} of $\mathcal{L}$ as \begin{align*}
        \eta := \inf_{f \in \ELL^2_0(\mu)\cap \mathcal D(\mathcal L), f\neq 0}\frac{\|\mathcal L f\|_{\ELL^2(\mu)}}{\|f\|_{\ELL^2(\mu)}}.
    \end{align*}
\end{definition}

\begin{remark}
    The singular value gap of generator $\mathcal{L}$ plays a special role in the study of Markov processes, see \cite{xu2026singular, eberle2026relaxation,chatterjee2025spectral}.  For example, \cite{huang2026bernstein} established a Bernstein-type inequality in terms of the singular value gap.  \cite{eberle2026relaxation} showed that the singular value gap of the two-point motion, which corresponds to the independent coupling of two copies of the process,  provides a lower bound on the usual spectral gap of the generator. 
\end{remark}

We now introduce two new functional inequalities, which further extend the iterated Poincar\'e and the weak flow Poincar\'e inequality.
\begin{definition}[Weak Iterated Poincar\'e inequality]
The semigroup $(P_t)_{t\geq 0}$ satisfies a \emph{weak iterated Poincar\'e inequality (WIPI)} with $\beta_{\mathrm I}: (0, \infty) \rightarrow [0, \infty)$ if for all $f \in \ELL^2_0(\mu) \cap \mathcal D(\mathcal L)$, $t\ge 0$ and $s>0$, 
    \begin{align}
    \label{eq: IWPI}
    \left\| P_tf\right\|_{\ELL^2(\mu)}^2  &\leq s \|\mathcal{L}P_tf\|^2_{\ELL^2(\mu)}  + \beta_{\mathrm I}(s)\|f\|^2_\osc,
\end{align}
where $\beta_{\mathrm I}$ is as in \Cref{def:WPI}.
\end{definition}

\begin{definition}[Weak Iterated Flow Poincar\'e inequality]
The semigroup $(P_t)_{t\geq 0}$ satisfies a \emph{weak iterated flow Poincar\'e inequality (WIFPI)} with $\BIF: (0, \infty) \rightarrow [0, \infty)$ and duration $T \in (0, \infty)$ if for all $f \in \ELL^2_0(\mu) \cap \mathcal D(\mathcal L)$  and $s>0$,
    \begin{align*}
        \int_0^T\left\| P_tf\right\|_{\ELL^2(\mu)}^2 \dif t &\leq s \int_0^T\|\mathcal{L}P_tf\|^2_{\ELL^2(\mu)} \dif t + \BIF(s)\|f\|^2_\osc,
    \end{align*}
    where $\BIF$ is as in \Cref{def:WPI}.
\end{definition}

\begin{remark}
    Since $\beta$ satisfies the conditions in \Cref{def:WPI}, we can also deduce a $K^*$ function and $F^{-1}$ formulation for these inequalities. However, the WIPI and WIFPI do not imply convergence of $(P_t)_{t\geq 0}$ in $\ELL^2$. We only use them as auxiliary inequalities in intermediate comparison steps.
\end{remark}

\section {Comparison of the WPIs}
\label{Sec: comparison}
In order to discuss the fastest possible acceleration due to lifting, we begin by describing a notion of an \textit{optimal} WPI. 

First, we define the optimal $\beta$ in the continuous time setting, building on the discrete time formulation in \cite{andrieu2026weak}. We start with the WPI and generalise the relative arguments into other types of weak flow Poincar\'e inequalities, namely WFPI and WIFPI. Note that the upper bounds for $\beta$ function differ. For the WPI and WIPI $\beta\leq 1/4$, whereas for WFPI and WIFPI with duration $T$, $\beta\leq T/4$. For greater generality, we consider $\beta\leq a$ for some $a>0$ in this section. 

\begin{definition}
    \label{def:betastar}
For any $s > 0$,  
\begin{align*}
    \beta^\star(s) :=\sup_{\|f\|^2_\osc=1, f \in \ELL^2_0(\mu)\cap \mathcal D(\mathcal{L})}\{\|f\|_{\ELL^2(\mu)}^2- s\cdot \left\langle f,-{\mathcal{L}} f\right\rangle_{\ELL^2(\mu)}  \} \lor 0,
\end{align*} 
noting that for $s > 0$, $\beta^\star(s) \leq a$.

Correspondingly, we define the optimal ${K}^*$ as follows:
\begin{align*}
    K^\star(v):=\sup_{u \ge  0}\{uv - u\beta^\star(1/u)\}.
\end{align*}
\end{definition}

Next, we consider the set of valid $\beta$ functions for $\mathcal{L}$.
\begin{definition}
    Suppose that the $\mu$-invariant semigroup $(P_t)_{t \geq 0}$ possesses some WPI for some functions $\beta$. We equip the set $\mathcal{B}$ with the partial order
     \begin{align*}
         \beta_1\leq \beta_2 \iff  \beta_1(s)\leq \beta_2(s), \quad \forall s>0.
     \end{align*}
     Let $\mathcal{B}^{\mathcal{L}}$ denote the set of valid $\beta$ functions for which $\mathcal{L}$ satisfies a $\beta$-WPI:
    \begin{align*}
        \mathcal{B}^\mathcal{L}=\{\beta \in \mathcal{B}: \left\| P_tf\right\|_{\ELL^2(\mu)}^2  \leq s \left\langle P_tf,-{\mathcal{L}} P_tf\right\rangle_{\ELL^2(\mu)}  + \beta(s)\|f\|^2_\osc,\\\forall s>0,  \quad \forall f \in \ELL^2_0(\mu)\cap \mathcal D(\mathcal{L})\}.
    \end{align*}
\end{definition}

\begin{lemma}
    $\mathcal{B}^\mathcal{L} \subset \mathcal{B}$, and $\beta^\star$ is the minimum element of $\mathcal{B}^\mathcal{L}$ with respect to the partial order.
\end{lemma}
\begin{proof}
    It follows from the definition of $\beta^\star$ that for any $\beta \in \mathcal{B}^\mathcal{L}$, we have $\beta^\star(s) \leq \beta(s)$ for all $s>0$. Hence $\beta^\star$ is the minimum element of $\mathcal{B}^\mathcal{L}$ with respect to the partial order.
\end{proof}

\begin{remark}
    In the SPI case, as in Remark~\ref{rmk:strongtoweak}, we can take $\beta(s)=\frac{1}{4}\cdot \mathbbm{1}_{\{s\leq \gap(\mathcal L)^{-1}\}}$. 
    Hence $\beta^\star(s)=0$ when $s\geq \gap(\mathcal L)^{-1}$.
\end{remark}

\begin{lemma}
    Suppose that the semigroup $(P_t)_{t\geq 0}$ possesses a $K^*$-WPI for some function $K^*$. Then $K^\star(\cdot) \geq K^* (\cdot)$ defines a $K^\star$-WPI.
\end{lemma}
\begin{proof}
    Let $K'(u):=\sup_{0\le v \le a}\{uv - K^*(v)\}$, and $\beta'(s):=s\cdot K'(1/s)\Leftrightarrow K'(u)=u\beta'(1/u)$. One can check $(P_t)_{t\geq 0}$ possesses a $\beta'$-WPI. By definition of $\beta^\star$, we have that $0\leq \beta^\star \leq \beta' $ pointwise. Hence we have that $0\leq K^* \leq K^\star $ pointwise.
\end{proof}

\begin{definition}
    We equip the set $\mathcal{C}$ with the partial order
     \begin{align*}
         K^*_1\leq K^*_2 \iff  K^*_1(v)\leq K^*_2(v), \quad \forall v \in[0, a].
     \end{align*}

    Let $\mathcal{C}^\mathcal{L}$ denote the set of valid $K^*$ functions for which $\mathcal{L}$ satisfies a $K^*$-WPI:
    \begin{align*}
        \mathcal{C}^\mathcal{L}=\left \{K^* \in \mathcal{C}: {K}^*(v)=\sup_{u\geq 0}\{uv - u\beta(1/u)\}, \beta \in \mathcal{B}^\mathcal{L}\right \}.
    \end{align*}
\end{definition}
\begin{lemma}
    $K^\star$ is the maximum element of $\mathcal{C}^\mathcal{L}$ with respect to the partial order.
\end{lemma}
\begin{proof}
    It follows from the definition of $K^\star$ that for any $K^* \in \mathcal{C}^\mathcal{L}$, we have $K^*(v) \leq K^\star(v)$ for all $v \in[0, a]$. Hence $K^\star$ is the maximum element of $\mathcal{C}^\mathcal{L}$ with respect to the partial order.
\end{proof}

Now, we consider the comparison of two processes and adopt the following notation throughout the remainder of this section. Let $(P_t)_{t\geq 0}$ and $(Q_t)_{t\geq 0}$ be two Markov semigroups on $\ELL^2(\mu_1)$ and $\ELL^2(\mu_2)$ with infinitesimal generators $(\mathcal L_1, \mathcal D(\mathcal L_1))$ and $(\mathcal L_2, \mathcal D(\mathcal L_2))$, respectively. Suppose that $\mathcal L_1$ satisfies a $\beta_1$-WPI and $\mathcal L_2$ a $\beta_2$-WPI respectively, and $K_1^*$ and $K_2^*$ correspond to $\beta_1$ and $\beta_2$, respectively. Let $\gamma_i = F^{-1}_i$  be the respective convergence bound functions as defined above in \Cref{thm:WPI_convergence}. {We denote the optimal $\beta$ for $\mathcal{L}_1$ and $\mathcal{L}_2$ by $\beta_1^\star$ and $\beta_2^\star$, respectively, and $K_1^\star$ and $K_2^\star$ be their corresponding $K^*$ functions. Let $\gamma_1^\star$ and $\gamma_2^\star$ be the respective convergence bound functions as in Theorem~\ref{thm:WPI_convergence}.}
\begin{theorem}
\label{thm: derict_compare_two}
If $\beta_2(\cdot) \geq \beta_1(\cdot)$, then $K^*_1 \geq K^*_2$ and $\gamma_2(\cdot)\geq \gamma_1(\cdot)$.
\end{theorem}

\begin{proof}
    \label{prf: thm: derict_compare_two}
    From the definitions, $K_1 \leq K_2$ and hence $K^*_1 \geq K^*_2$. As a result, $F_1 \leq F_2$ and consequently $\gamma_1 := F^{-1}_1 \leq F^{-1}_2 =: \gamma_2$.
\end{proof}

\begin{remark}
    Once we obtain a lower bound for $\beta^\star$ or an upper bound for $K^\star$, these then imply a lower bound on the fastest rate of convergence possible.  This should be distinguished from a lower bound on the convergence error over all the functions under consideration.
    
    For example, \Cref{eq:convergence rate} is equivalent to 
    \begin{align*}
    \sup_{\|f\|^2_\osc=1, f \in \ELL^2_0(\mu)\cap \mathcal D(\mathcal L)} \|P_tf \|^2_{\ELL^2(\mu)} \leq  \gamma(t).
    \end{align*}
    Hence, 
    $\gamma^\star(t):=  \sup_{\|f\|^2_\osc=1, f \in \ELL^2_0(\mu)\cap \mathcal D(\mathcal L)} \|P_tf \|^2_{\ELL^2(\mu)}$ represents the optimal convergence bound. 
    Our aim is to lower bound $\gamma^\star $ rather than establish a lower bound on the left-hand side $ \|P_tf \|^2_{\ELL^2(\mu)}$ for any particular $f$.
\end{remark}

Based on this comparison idea, we present the following theorem over optimal WPIs.

\begin{theorem}
\label{thm: optimal_comparison}
    Consider two semigroups with infinitesimal generators $(\mathcal L_1, \mathcal D(\mathcal L_1))$ and $(\mathcal L_2, \mathcal D(\mathcal L_2))$ satisfying WPIs as described above. 
    Suppose the processes are such that whenever  
    $\mathcal L_1$ satisfies a $\beta_1$-WPI, then $\mathcal L_2$ also satisfies a $\beta_1$-WPI. 
    Then $\beta^\star_2\leq \beta^\star_1$, $K^\star_2\geq K^\star_1$ and the corresponding respective convergence bounds satisfy $\gamma_1^\star(\cdot)\geq \gamma_2^\star(\cdot)$.
\end{theorem}
\begin{proof}
    \label{prf: thm: optimal_comparison}
    For all $\beta_1 \in \mathcal{B}^{\mathcal{L}_1}$, we have $\beta_1 \in \mathcal{B}^{\mathcal{L}_2}$, since $\mathcal L_2$ also satisfies a $\beta_1$-WPI. Then $\beta^\star_1(s)\in \mathcal{B}^{\mathcal{L}_2}$. Hence $\beta^\star_2(s) \leq \beta^\star_1(s)$ for all $s>0$ as $\beta^\star_2$ is the minimum element of $\mathcal{B}^{\mathcal{L}_2}$. Next, by \Cref{thm: derict_compare_two}, we have that $K^\star_2(v) \geq K^\star_1(v)$ for all $v \in[0, a]$, so $\gamma_1^\star(\cdot)\geq \gamma_2^\star(\cdot)$.
\end{proof}

Now we consider a more complex comparison scenario involving linear transforms.

\begin{theorem}
\label{thm: linear T}
    Consider two semigroups with infinitesimal generators $(\mathcal L_1, \mathcal D(\mathcal L_1))$ and $(\mathcal L_2, \mathcal D(\mathcal L_2))$ satisfying WPIs as described above. If for any $K^*_1(v) \in \mathcal{C}^{\mathcal{L}_1}$, $\mathcal L_2$ also satisfies a $ \left(\frac{1}{c} \cdot K^*_1\right)$-WPI, then $K^\star_2(v)\geq \frac{1}{c} \cdot K^\star_1(v)$, so $K^\star_1(v)\leq c \cdot K^\star_2(v)$. 
Hence, the corresponding convergence bounds for $\mathcal L_1$ satisfy $\gamma_1^\star(\cdot)\geq \gamma^\star(\cdot)$, where $\gamma^\star(\cdot)$ corresponds to $c \cdot  K^\star_2$.
\end{theorem}

\begin{proof}
\label{prf: thm: linear T}
    For all $K^*_1 \in \mathcal{C}^{\mathcal{L}_1}$, we have $K^*_1/c \in \mathcal{C}^{\mathcal{L}_2}$, since $\mathcal L_2$ also satisfies a $ K^*_1/c$-WPI. Then $K^\star_1/c\in \mathcal{C}^{\mathcal{L}_2}$. Hence $K^\star_2 \geq K^\star_1/c$ as $K^\star_2$ is the maximum element of $\mathcal{C}^{\mathcal{L}_2}$. Next, by \Cref{thm: derict_compare_two}, we have that $c\cdot K^\star_2(v) \geq K^\star_1(v)$ for all $v \in[0, a]$, $\gamma_1^\star(\cdot)\geq \gamma^\star(\cdot)$, where $\gamma^\star(\cdot)$ corresponds to $c \cdot  K^\star_2$.
\end{proof}

\begin{example}
\label{lem:compare dirichlet form}
    {Suppose $\mu_1 =\mu_2=\mu$, and for some $c>0$, if $\left\langle f,-{\mathcal{L}_1} f\right\rangle_{\ELL^2(\mu)} \leq c\left\langle f,-{\mathcal{L}_2} f\right\rangle_{\ELL^2(\mu)}$, $\mathcal L_1$ satisfies a $\beta^\star_1$-WPI, and $\mathcal L_2$ satisfies a $\beta^\star_2$-WPI, then $\beta^\star_1(s) \geq \beta^\star_2(c \cdot s)$, $K^\star_1(v) \leq c \cdot  K^\star_2(v)$, and $\gamma_1^\star(\cdot)\geq \gamma_2^\star(c\cdot )$.}
\end{example}
\begin{proof}
    From the ordering of Dirichlet forms, we have for any $f \in \mathrm L^2(\mu)$ that  
    \begin{align*}
        \|f\|^2_{\ELL^2(\mu)} - s \cdot \left\langle f,-{\mathcal{L}_1} f\right\rangle_{\ELL^2(\mu)}\geq \|f\|^2_{\ELL^2(\mu)} - s \cdot c\left\langle f,-{\mathcal{L}_2} f\right\rangle_{\ELL^2(\mu)},
    \end{align*}
    so from \Cref{def:betastar} we deduce $\beta^\star_1(s) \geq \beta^\star_2(c\cdot s)$. From Lemma 14 in \cite{andrieu2022comparison}, we then conclude that $K^\star_1(v) \leq c \cdot  K^\star_2(v)$ and $\gamma_1^\star(\cdot)=F^{-1}_1(\cdot)\geq F^{-1}_2(c\cdot )=\gamma_2^\star(c\cdot )$.
\end{proof}

\begin{remark}
\Cref{lem:compare dirichlet form} is a special case of \Cref{thm: linear T}. Since in the scenario of \Cref{lem:compare dirichlet form}, $\left\langle f,-{\mathcal{L}_1} f\right\rangle_{\ELL^2(\mu)} \leq c\left\langle f,-{\mathcal{L}_2} f\right\rangle_{\ELL^2(\mu)}$ means if $\mathcal{L}_1$ admits $\beta_1$-WPI, then
\begin{align*}
    \left\|f\right\|_{\ELL^2(\mu)}^2  &\leq s \left\langle f,-{\mathcal{L}_2} f\right\rangle_{\ELL^2(\mu)}  + \beta_1(s/c)\|f\|^2_\osc,\quad\forall s>0.
\end{align*}
By Lemma 14 in \cite{andrieu2022comparison}, it gives for all $K^*_1\in \mathcal{C}^{\mathcal{L}_1}$, $L_2$ also satisfies the $ \left(\frac{1}{c} \cdot K^*_1\right)$-WPI.
\end{remark}

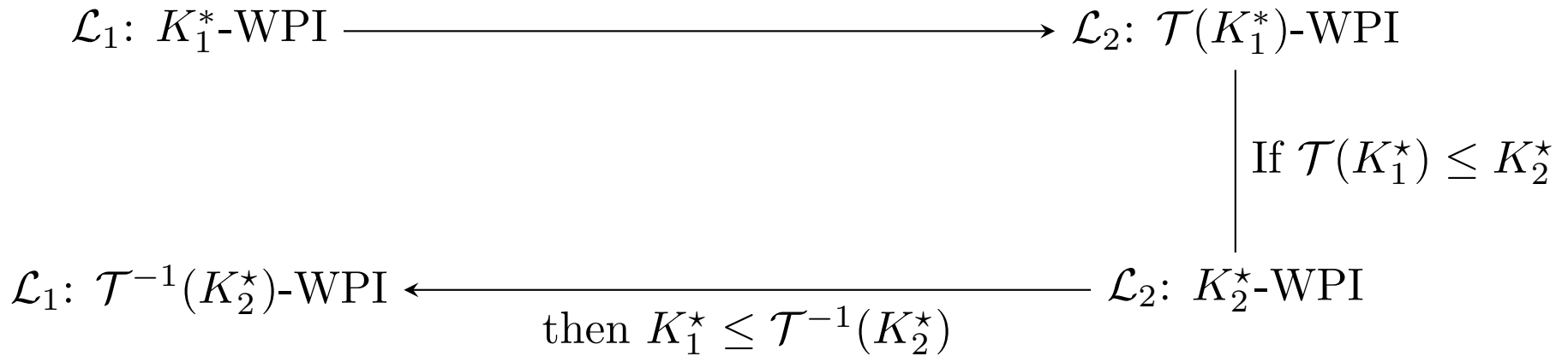
\begin{figure}[H] 
    \centering 
    \begin{tikzpicture}[>=stealth]
        \node (P) at (2,2) {$\mathcal{L}_1$: $K_1^*$-WPI};
        \node (P^*) at (10,2) {$\mathcal{L}_2$: $\mathcal{T}(K_1^*)$-WPI };
        \node (P_{(2)}^*) at (10,0) {$\mathcal{L}_2$:  $K^\star_2$-WPI };
        \node (P_{(2)}) at (2,0) {$\mathcal{L}_1$: $\mathcal{T}^{-1}(K_2^\star)$-WPI};

        \draw [->] (P) --  (P^*); 
        \draw [-] (P^*) -- node[right] { If $\mathcal{T}(K_1^\star)\leq K_2^\star$}
        (P_{(2)}^*);
        \draw [->] (P_{(2)}^*) -- node[below] {then $K_1^\star\leq \mathcal{T}^{-1}(K_2^\star)$} (P_{(2)});
        
    \end{tikzpicture}
    \caption{{General Functional $\mathcal{T}$}} 
    \label{fig: T} 
\end{figure}

\Cref{thm: linear T} shows that after a linear transformation, the properties of $K^*$ are preserved. It is natural to extend this comparison under broader classes of transformations. To be more precise, if $\mathcal{T}$ is a transformation from $\mathcal{C}$ to $\mathcal{C}$, and $\mathcal{T}$ is a bijection in $\mathcal{C}$, then given a $K^*\in \mathcal{C}$, $\mathcal{T}(K^*)\in \mathcal{C}$, and $\mathcal{T}^{-1}(K^*)\in \mathcal{C}$. Therefore, $\mathcal{T}(K^*)$ and $\mathcal{T}^{-1}(K^*)$ are both admissible as $K^*$-functions to give an $\mathcal{T}(K^*)$-WPI or $\mathcal{T}^{-1}(K^*)$-WPI, respectively. Meanwhile, if $\mathcal{T}$ preserves the partial order of $K^*$ (as the linear transformation does), we could obtain an upper bound for the optimal $K^\star$ via different inequalities, see in \Cref{fig: T}.

{However, for some transformations, $\mathcal{T}$ and $\mathcal{T}^{-1}$ preserve the admissibility of $K^*$ functions only on a subset of $\mathcal{C}$.} Hence, this motivates the following assumption to specify that under these conditions the transformation preserves the class of admissible $K^*$-functions.

let $\mathcal{T}:\mathcal{C}\to \mathbb{R}^{[0, a]}$ be a functional, where $\mathbb{R}^{[0, a]}$ denotes the set of functions $g:[0, a]\to \mathbb R$, $\mathcal{C}$ is the collection of all $K^*$ functions \eqref{eq:C_def}. 

{Firstly, we introduce a more tractable way to describe the properties of $K^*$ functions based on \cite[Proposition 4]{andrieu2026weak}. The proof of the following \Cref{thm:C} is in \Cref{prf:thm:C}.
\begin{lemma}
\label{thm:C}
With $\mathcal C$ defined as in \eqref{eq:C_def}, we define 
\begin{align*}
    \mathcal C_0:=\left\{K^*:[0,a)\rightarrow [0, \infty) \mid K^* \text{is nonnegative, convex, strictly increasing and } K^*(0)=0\right\}.
\end{align*}
Then $\mathcal{C}_0\subset \mathcal C|_{[0,a)}$, where $\mathcal C|_{[0,a)}$ denotes the class of functions in $\mathcal C$ restricted to the domain $[0,a)$.
\end{lemma}}

\begin{remark}
    Since $\beta$ can be chosen such that $\beta(s)\leq a$ for all $s>0$, $K^*(v)=\infty$ as $v>a$ where $K^*$ corresponds to $\beta$ defined as \Cref{thm:K*}. Therefore, it suffices to work with $\mathcal C|_{[0,a)}$.
\end{remark}

\begin{assumption}
\label{asp: T}
For a fixed set $\mathcal{C'}\subset\mathcal{C}$, let $\mathcal{\overline T}:=\mathcal{T}|_{\mathcal C'}$ such that the following holds: 
\begin{enumerate}[(A)]

    \item {$\mathcal{T}(\mathcal{C}) \subset \mathcal{C}_0$}.
    \item $\mathcal{\overline T}:\mathcal C' \rightarrow \mathcal C$ is an injection.
    \item $\forall K_1^*, K_2^* \in \mathcal{C'}$, $K_1^* \leq K_2^*\iff \mathcal{\overline T}(K_1^*)\leq \mathcal{\overline T}(K_2^*)$.
\end{enumerate}
\end{assumption}

\begin{example}
\label{emp:T}
    \Cref{asp: T} ensures that, for a given class $\mathcal C'$ of $K^*$ functions, applying the functional $\mathcal{T}$ produces another valid $K^*$ function. Examples include:
    \begin{itemize}
        \item Consider $\mathcal{T}_1:K^*(v) \mapsto cK^*(v)$ as in \Cref{thm: linear T}, where $c>0$, then it is natural to take $\mathcal{C'}=\mathcal{C}$,  and $(\mathcal C,\mathcal{T}_1)$ automatically satisfies \Cref{asp: T} by Theorem~\ref{thm: linear T}.
        \item Suppose that 
        $$\mathcal{T}_2:K^* \mapsto 
        \begin{cases}
            (K^*(v))^2/v, &  v>0;\\
            0, & v=0.
        \end{cases}$$ 
        One can check \Cref{asp: T} holds in these three cases:
        \begin{itemize}
            \item $\mathcal C_1'=\{K^*(v)=Cv: C>0, 0\leq v < a \}$;
            \item $\mathcal C_2'=\{K^*(v)=Cv^{1+c_1^{-1}}: c_1>0,C>0, 0\leq v < a \}$;
            \item $\mathcal C_3'=\{K^*(v)=Cv(\log(1/v))^{-\frac{1}{\eta}}: C>0, \eta>0, 0\leq v < a \}$.
        \end{itemize}
    \end{itemize}
\end{example} 

\begin{proof}[Proof of \Cref{emp:T}]
\label{prf:emp:T}
    {We only check (A) in \Cref{asp: T} for $\mathcal{T}_2$.}

    {It is clear that $\mathcal{T}_2(K^*)$ is nonnegative when $v\geq 0$, and $\mathcal{T}_2(K^*)(0)=0$.}
    
    {When $v>0$, we have
    \begin{align*}
        \mathcal{T}_2(K^*)(v)'=\left(\frac{K^*(v)^2}{v}\right)'=\frac{K^*(v)}{v^2}\left(2vK^*(v)'-K^*(v)\right)> 0,
    \end{align*}
    since $vK^*(v)'> K^*(v)$ by the convexity of $K^*$ and $K^*(0)=0$. Hence $\mathcal{T}_2(K^*)$ is strictly increasing as $v>0$.}

    {As $K^*$ is convex, $\mathcal{T}_2(K^*)$ is also convex since
    \begin{align*}
        \mathcal{T}_2(K^*)(v)''=\left(\frac{K^*(v)^2}{v}\right)''=\frac{2}{v}\left( K^*(v)K^{*}(v)''+\left(K^{*}(v)'-\frac{K^*(v)}{v}\right)^2\right)\geq 0.
    \end{align*}}
\end{proof}

\begin{remark}
    As shown in \cite[Lemma 15, Lemma 16]{andrieu2022comparison}, if $K^* \in \mathcal C'_2$, then we have polynomial convergence $\gamma(t) \leq C't^{-c_1}$; if $K^* \in \mathcal C'_3$, then we have stretched exponential convergence $\gamma(t) \leq C'' \exp \left(-\left(C\frac{1+\eta}{\eta}t\right)^{\eta/1+\eta}\right).$
\end{remark}

\begin{lemma}
    Under \Cref{asp: T}, $\forall K_1^*, K_2^* \in \mathrm{Range}(\mathcal{\overline T})$, $K_1^* \leq K_2^*\iff \mathcal{\overline T}^{-1}(K_1^*)\leq \mathcal{\overline T}^{-1}(K_2^*)$.
\end{lemma}
\begin{proof}
As $\mathcal{\overline T}^{-1}(K_1^*), \mathcal{\overline T}^{-1}(K_2^*) \in \mathcal{C}'$, $ \mathcal{\overline T}^{-1}(K_1^*)\leq \mathcal{\overline T}^{-1}(K_2^*)\iff  \mathcal{\overline T}(\mathcal{\overline T}^{-1}(K_1^*))\leq \mathcal{\overline T}(\mathcal{\overline T}^{-1}(K_2^*))$.
\end{proof}

\begin{corollary}
\label{thm: comparisonT}
{Suppose there exist $\mathcal C' \subset \mathcal C$ and $\mathcal{T}:\mathcal{C}\to \mathbb{R}^{[0, a]}$ such that  $(\mathcal C', \mathcal{T})$ satisfies \Cref{asp: T}. 
Consider two semigroups with infinitesimal generators $(\mathcal L_1, \mathcal D(\mathcal L_1))$ and $(\mathcal L_2, \mathcal D(\mathcal L_2))$ satisfying WPIs.} 

{Suppose that for any $K^*_1 \in \mathcal{C}^{\mathcal{L}_1}$, $\mathcal L_2$ also satisfies a $\mathcal{\overline T}(K^*_1)$-WPI, and that the maximal $K^\star_2\in \mathcal{\overline T}(\mathcal{C'})$. 
Then, the corresponding convergence bounds for $\mathcal L_1$ satisfy $\gamma_1^\star(\cdot)\geq \gamma^\star(\cdot)$, where $\gamma^\star(\cdot)$ corresponds to $\mathcal{\overline T}^{-1}( K^\star_2)$.}
\end{corollary}
\begin{proof}
    Under \Cref{asp: T}, we note that
    for all $K^*_1 \in \mathcal{C}^{\mathcal{L}_1}$, $\mathcal{\overline T}(K^*_1) \in \mathcal{C}_0 \subset \mathcal C$, which means $\mathcal{\overline T}(K^*_1)$ is valid to define a WPI. Then, we have $K^*_1 \in \mathcal{C}^{\mathcal{L}_2}$, since $\mathcal L_2$ also satisfies a $ \mathcal{\overline T}(K^*_1)$-WPI. 
    Then $\mathcal{\overline T}(K^*_1)\in \mathcal{C}^{\mathcal{L}_2}$. Hence $K^\star_2 \geq \mathcal{\overline T}(K^*_1)$ as $K^\star_2$ is the maximum element of $\mathcal{C}^{\mathcal{L}_2}$. Since we are assuming $K^\star_2 \in \mathcal{\overline T}(\mathcal C')$, $\mathcal{\overline T}^{-1} (K^\star_2) \in \mathcal C'$. Next, by \Cref{asp: T}, we have that $\mathcal{\overline T}^{-1} (K^\star_2) \geq K^\star_1(v)$ for all $v \in[0, a]$, so $\gamma_1^\star(\cdot)\geq \gamma^\star(\cdot)$, where $\gamma^\star(\cdot)$ corresponds to $\mathcal{\overline T}^{-1} (K^\star_2)$.
\end{proof}

\begin{remark}
    All the definitions and theorems can be generalised into other types of weak Poincar\'e inequalities, namely WFPI and WIFPI. For example, we denote $\mathcal{B}_F^{\mathcal{L}}$ as the set of functions $\BF$ for which $\mathcal{L}$ satisfies a $\BF$-WFPI, and by $\mathcal{C}_F^{\mathcal{L}}$ the corresponding set of functions $\KF$. Similarly, we denote the corresponding sets for WIFPI by $\mathcal{B}_{IF}^{\mathcal{L}}$ and $\mathcal{C}_{IF}^{\mathcal{L}}$.
\end{remark}

\begin{remark}
    \Cref{thm: comparisonT} shows the comparison WPI between two different semigroups. The result can be easily generalised to the comparisons between other types of inequalities, as the proof only involves the corresponding sets of admissible $\beta$ or $K^*$ functions.
\end{remark}

\section{Main result}
\label{Sec: Main result}
Our aim is to show the following: Firstly, if $\mathcal{L}$ satisfies some WPI, it satisfies an optimal $(\beta^\star, K^\star)$-WPI. Then, through a sequence of intermediate comparisons, we obtain a \textit{lower bound} on the convergence rate of the Markov process generated by $\hat{\mathcal{L}}$. Throughout the rest of the paper, we will use the ‘hat’ notation to refer
to a lifted chain; for example, $\widehat{K^*_\mathrm F}$ is the WFPI $K^*$ function of a lifted chain $(\hat P_t)_{t\geq 0}$, where the subscript indicates the different forms of inequality.

\begin{theorem}
\label{thm: main}
Assume that $\hat{\mathcal{L}}$ is a second-order lift of $\mathcal{L}$, and that
$\mathcal{L}$ satisfies a WPI. Without loss of generality, suppose that
$\mathcal{L}$ satisfies a $\beta^\star$-WPI and that $\hat{\mathcal{L}}$
satisfies some $\LKF$-WFPI with duration $T \in (0, \infty)$. 

We write $\mathcal{C}^{\mathcal{L}}$ and $\mathcal{C}_{\mathrm F}^{\mathcal{\hat L}}$ for the valid set of WPI $K^*$-functions for $\mathcal{L}$ and WFPI $\LKF$-functions for $\mathcal{\hat{L}}$, respectively. Assume that there exists $\mathcal{C'}$ such that $\mathcal{C'}\subset \mathcal{C}$. Define
        \begin{align*}
            \mathcal{T}: K^*(v) &\mapsto \frac{(K^*(v))^2}{v}, \quad v>0;\\
            \mathcal{T}(0)&=0.
        \end{align*}
        Suppose $(\mathcal C', \mathcal{T})$ satisfies \Cref{asp: T}. Then the following statements hold.
    \begin{enumerate}
        \item There exists some $\LBI \in \mathcal{B}$ such that $\mathcal{\hat L}$
    satisfies a $\LBI$-WIPI. Moreover, there exists some
    $\LBIF \in \mathcal{B}$ such that $\hat{\mathcal{L}}$
    satisfies a $\LBIF$-WIFPI.
        \item  Then for any $\LKF$-WFPI satisfied by $\hat{\mathcal{L}}$, we have
    \begin{align*}
    \LKF(v) &\leq \sqrt{v \cdot \LKIF(v)},
    \end{align*}
    where $\LKIF(v):= \sup_{u\geq 0}\{uv - u \LBIF (1/u)\}$, and $\LBIF(s) \geq \frac{T}{2}\beta^\star(4T^2+4s)$.
    
    {When $\LKIF(v)\in\mathcal{\overline T}(\mathcal C')$, the corresponding convergence bounds for $\hat{\mathcal L}$ satisfy $\gamma(\cdot)\geq \hat \gamma_{\mathrm{IF}}(\cdot)$, where $\hat \gamma_{\mathrm{IF}}$ corresponds to $\sqrt{v \cdot \LKIF(v)}$.}
    \end{enumerate}
\end{theorem}

\begin{remark}
    This gives a lower bound on the convergence behavior of $\hat{\mathcal{L}}$, based on the WPI of $\mathcal{L}$. This limts the acceleration deduced by lifting.
\end{remark}

For proving \Cref{thm: main}, here are the comparison ideas:
\begin{itemize}
    \item  If $\mathcal{\hat L}$ satisfies a $\LKF$-WFPI, then $\mathcal{\hat L}$ satisfies a $\LKIF$-WIFPI, where $\LKIF$ is the function $v\mapsto (\LKF(v))^2/v$.
    \item  If $\mathcal{\hat L}$ satisfies a $\LBIF$-WIFPI, then $\mathcal{\hat L}$ satisfies a $\LBI$-WIPI, where $\LBI=\frac{2}{T}\LBIF\left(\frac{s'-2T^2}{2}\right)$, for $s>2T^2$.
    \item If $\mathcal{\hat L}$ satisfies a $\LBI$-WIPI, then $\mathcal{L}$ satisfies a $\beta$-WPI, where ${\beta}(s) = \LBI(s/2)$.
\end{itemize}

\begin{figure}[H] 
    \centering 
    \begin{tikzpicture}[>=stealth]
        \node (P) at (2,2) {$\mathcal L$: $\beta^\star$-WPI };
        \node (P^*) at (10,2) { $\mathcal{\hat L}$: $\OLBI$-WIPI };
        \node (P_{(2)}^*) at (10,0) {$\hat{\mathcal L}$:  $\OLBIF$-WIFPI};
        \node (P_{(2)}) at (2,0) {$\hat{\mathcal L}$: $\OLKF$-WFPI};

        \draw [->] (P) -- node[above] {\Cref{thm:reversible}} node[below] {$\OLBI(s) \geq \beta^\star(2s)$}  (P^*); 
        \draw [->] (P^*) -- node[right] {$\OLBIF\geq\frac{T}{2}\OLBI(2T^2+2s)$}
        node[left] {\Cref{thm:lift}}  (P_{(2)}^*);
        \draw [->] (P_{(2)}^*) -- node[above] {\Cref{thm:order}}
        node[below] {$\OLKF(v) \leq \sqrt{\OLKIF(v) \cdot v}$} (P_{(2)});
        
    \end{tikzpicture}
    \caption{Comparison among inequalities.} 
    \label{fig:comparison} 
\end{figure}
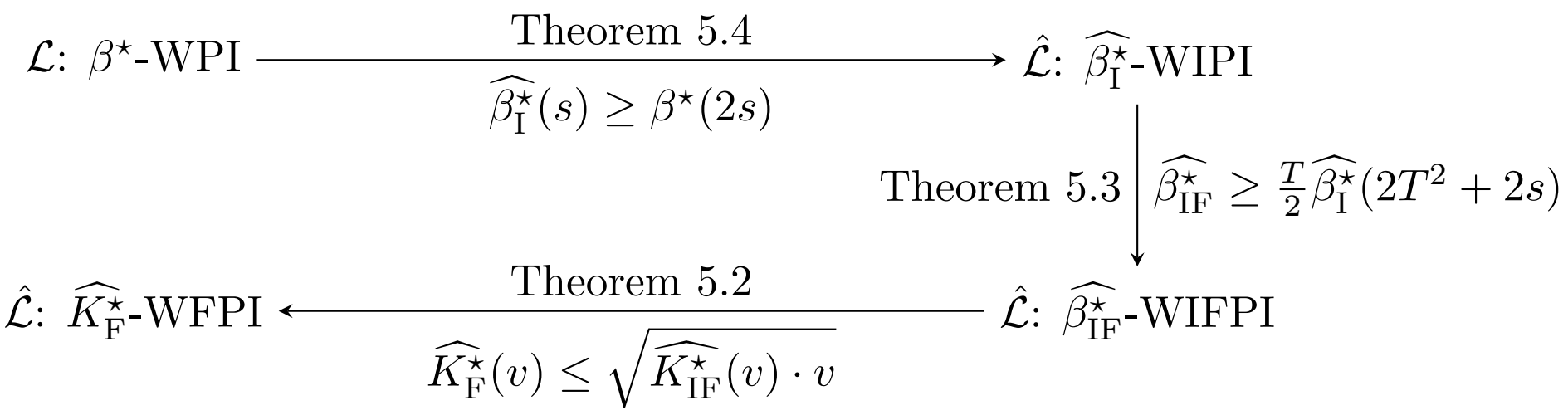

Therefore, we can derive a lower bound for the optimal WFPI convergence behavior of $\hat{\mathcal L}$ by \Cref{thm: optimal_comparison} and \Cref{thm: comparisonT}. These comparisons are summarized in the schematic diagram presented in \Cref{fig:comparison}. The complete theorem is shown in \Cref{thm: main}.

The proof of \Cref{thm: main} is deferred to later in \Cref{prf: thm: main} after we established the intermediate comparison steps.

\begin{theorem}
\label{thm:order} 
Consider the functional $\mathcal{T}: \mathcal A \to \mathbb R^{[0,a]}$ given by 
\begin{equation}
    \begin{split}
        \mathcal{T}: K^* &\mapsto \widehat{K^*},\\
        \widehat{K^*}&:=\left[ v\mapsto \frac {(K^*(v))^2} v,v>0,\quad  \widehat{K^*}(0)=0\right ].
    \end{split}
\end{equation} 
 
Suppose $(\hat P_t)_{t\geq 0}$ satisfies WFPI with $\LKF\in \mathcal{C}_F^{\mathcal{\hat L}}$ and duration $T \in (0, \infty)$, and there exists $\mathcal{C'}$ such that $\mathcal{C'}\subset \mathcal{C}$ and $(\mathcal{C'}, \mathcal{T})$ satisfies \Cref{asp: T},  then it satisfies a WIFPI with $ \LKIF(v)=(\LKF(v))^2/v$ and duration $T \in (0, \infty)$.
\end{theorem}

\begin{proof} 
For all $f \in \ELL^2_0(\mu)\cap \mathcal D(\hat{\mathcal L})$,
since $\left\langle \hat P_tf,-\hat{\mathcal{L}}\hat P_tf\right\rangle_{\ELL^2(\hat\mu)} \leq \|\hat P_tf\|_{\ELL^2(\hat\mu)}\|\hat{\mathcal{L}}\hat P_tf\|_{\ELL^2(\hat\mu)}$,
    we have
    
    \begin{align*}
        \LKF\left(\frac{    \int_0^T\left\| \hat{P_t}f\right\|_{\ELL^2(\hat\mu)}^2 \dif t}{\|f\|^2_\osc}\right)&\leq \frac{\int_0^T\left\langle \hat{P_t}f,-\hat{\mathcal{L}}\hat{P_t}f\right\rangle_{\ELL^2(\hat\mu)} \dif t}{\|f\|^2_\osc}\\
        &\leq \frac{\int_0^T \|\hat{\mathcal{L}}\hat{P_t}f\|_{\ELL^2(\hat\mu)}\|\hat{P_t}f\|_{\ELL^2(\hat\mu)} \dif t}{\|f\|^2_\osc}\\
        &\leq \frac{(\int_0^T \|\hat{\mathcal{L}}\hat{P_t}f\|^2_{\ELL^2(\hat\mu)} \dif t)^{1/2}(\int_0^T \|\hat{P_t}f\|^2_{\ELL^2(\hat\mu)} \dif t)^{1/2}}{\|f\|^2_\osc}\\
        &=\left(\frac{    \int_0^T\left\| \hat{P_t}f\right\|_{\ELL^2(\hat\mu)}^2 \dif t }{\|f\|^2_\osc}\right)^{1/2}\left(\frac{    \int_0^T\left\| \hat{\mathcal{L}} \hat{P_t}f\right\|_{\ELL^2(\hat\mu)}^2 \dif t }{\|f\|^2_\osc}\right)^{1/2}.
    \end{align*}
    Hence 
    \begin{align*}
        \left(\LKF\left(\frac{    \int_0^T\left\| \hat{P_t}f\right\|_{\ELL^2(\hat\mu)}^2 \dif t}{\|f\|^2_\osc}\right)\right)^2\leq \left(\frac{    \int_0^T\left\| \hat{P_t}f\right\|_{\ELL^2(\hat\mu)}^2 \dif t }{\|f\|^2_\osc}\right)\left(\frac{    \int_0^T\left\| \hat{\mathcal{L}} \hat{P_t}f\right\|_{\ELL^2(\hat\mu)}^2 \dif t }{\|f\|^2_\osc}\right).
    \end{align*}
    Let $\LKIF(v)=(\LKF(v))^2/v$, we obtain 
    \begin{align*}
        \LKIF\left(\frac{    \int_0^T\left\| \hat{P_t}f\right\|_{\ELL^2(\hat\mu)}^2 \dif t}{\|f\|^2_\osc}\right) \leq \frac{    \int_0^T\left\| \hat{\mathcal{L}} \hat{P_t}f\right\|_{\ELL^2(\hat\mu)}^2 \dif t }{\|f\|^2_\osc}.
    \end{align*}
    
    {Therefore, $\LKIF \in \mathcal C$ by the proof of \Cref{emp:T}}, which also shows the existence of a WIFPI for $\mathcal{\hat L}$ if it admits a WFPI.
\end{proof}

\begin{corollary}
\label{cor:order}
    Under the same condition of \Cref{thm:order}, suppose $ \OLKIF \in \mathcal{\overline T}(\mathcal{C'})$, then $\OLKF(v)\leq \sqrt{\OLKIF(v) \cdot v}$.
\end{corollary}
\begin{proof}
    Combining \Cref{thm:order} and \Cref{thm: comparisonT}, the optimal $\OKIF $ of WIFPI cannot be worse than $(\OKF(v))^2/v$. Thus if $\hat{\mathcal{L}}$ satisfies WIFPI with the optimal $\OLKIF(v)$ and it is in the range of $\mathcal{\overline T}$, then the optimal $\OKF$ of WFPI cannot be better than $\sqrt{\OLKIF(v)\cdot v}$.
\end{proof}

\begin{theorem}
\label{thm:lift}
    ection{If $(\hat P_t)_{t\geq 0}$ satisfies a $\beta$-WIFPI with $\LBIF$ and duration $T \in (0, \infty)$, then it satisfies WIPI with $\LBI(s)=\frac{2}{T}\LBIF\left(\frac{s-2T^2}{2}\right)$ for $s\ge 2T^2$.}
\end{theorem}
\begin{proof}
ection{
For $f \in \ELL^2_0(\hat \mu)\cap \mathcal D(\mathcal {\hat L})$, one can show that $\|\hat P_{t} f\|_{\ELL^2(\hat\mu)}^2$ and $\|\mathcal {\hat L}\hat P_{t} f\|_{\ELL^2(\hat\mu)}^2$ are both decreasing since 
    \begin{align*}
        \frac{ \dif \|\hat P_{t} f\|_{\ELL^2(\hat\mu)}^2}{\dif t}= -2 \left\langle \hat P_{t} f,
-\mathcal {\hat L}\hat P_{t} f\right\rangle \leq 0,
    \end{align*}
    and 
    \begin{align*}
\dif \|\mathcal {\hat L}\hat P_{t+h} f\|_{\ELL^2(\hat\mu)}^2= \|\hat P_{h} \mathcal {\hat L}\hat P_{t} f\|_{\ELL^2(\hat\mu)}^2 \leq \dif \|\mathcal {\hat L}\hat P_{t} f\|_{\ELL^2(\hat\mu)}^2,
    \end{align*}
for any $h>0$, since $\mathcal {\hat L}\hat P_{t} f=\hat P_{t}\mathcal {\hat L} f$ and $f$ is real-valued.
}

ection{
We apply the WIFPI to $\hat P_{t} f$, to obtain
    \begin{align*}
        \int_t^{t+T} \|\hat P_{u} f\|_{\ELL^2(\hat\mu)}^2 \dif u &\leq s \int_t^{T+t}\|\mathcal {\hat L}\hat P_{u} f\|^2_{\ELL^2(\hat\mu)}\dif u + \LBIF(s)\|\hat P_{t} f\|^2_\osc,\\
         \Rightarrow T \|\hat P_{T+t} f\|_{\ELL^2(\hat\mu)}^2  &\leq T s \|\mathcal {\hat L}\hat P_{t} f\|^2_{\ELL^2(\hat\mu)}  + \LBIF(s)\|\hat P_{t} f\|^2_\osc.
    \end{align*}
Following the proof of Theorem 7 in \cite{eberle2026relaxation}, note that for any $f\in \ELL^2_0(\hat \mu)\cap \mathcal D(\mathcal {\hat L})$, we have $\hat P_tf-f= \int_0^t \hat P_u\mathcal{\hat L}f \dif u$, so that 
\begin{align*}
    \|\hat P_tf-f\|_{\ELL^2(\hat\mu)}\leq \int_0^t \|\hat P_u\mathcal{\hat L}f \|_{\ELL^2(\hat\mu)}\dif u\leq t\|\mathcal{\hat L}f\|_{\ELL^2(\hat\mu)}.
\end{align*}
This yields
\begin{align*}
    \|f\|_{\ELL^2(\hat\mu)}^2 \leq 2\|\hat P_tf\|_{\ELL^2(\hat\mu)}^2 + 2t^2 \|\mathcal{\hat L} f\|_{\ELL^2(\hat\mu)}^2.
\end{align*}
 Hence, we can obtain
    \begin{align*}
         \|\hat P_tf\|_{\ELL^2(\hat\mu)}^2 &\leq 2\|\hat P_{t+T}f\|_{\ELL^2(\hat\mu)}^2 + 2T^2 \|\mathcal{\hat L} \hat P_tf\|_{\ELL^2(\hat\mu)}^2\\
        \Rightarrow \|\hat P_tf\|_{\ELL^2(\hat\mu)}^2 &\leq (2T^2+2s)\|\mathcal{\hat L} \hat P_tf\|_{\ELL^2(\hat\mu)}^2 + \frac{2}{T}\LBIF(s)\|P_{t} \tilde f\|^2_\osc.
    \end{align*}
    Let $s'=2T^2+2s$, 
    \begin{align*}
        \|\hat P_tf\|_{\ELL^2(\hat\mu)}^2 &\leq s'\|\mathcal{\hat L} \hat P_tf\|_{\ELL^2(\hat\mu)}^2 + \frac{2}{T}\LBIF\left(\frac{s'-2T^2}{2}\right)\|f\|^2_\osc.
    \end{align*}
    Then, we choose $\LBI(s)=1/4$ for $0 < s \leq 2T^2$, and $\LBI(s)=\min\left\{\frac{1}{4}, \frac{2}{T}\LBIF\left(\frac{s'-2T^2}{2}\right)\right\}$ for $ s > 2T^2$.
}
\end{proof}

\Cref{thm:lift} implies if $\mathcal{\hat L}$ satisfies a WIPI with the optimal $\OLBI(s)$, then the optimal $\OLBIF(s)$ for a WIFPI for $\hat{\mathcal{L}}$ cannot be better than $\frac{T}{2}\OLBI(2T^2+2s)$. 

\begin{theorem}
\label{thm:reversible}
    ection{If $\mathcal {\hat L}$ satisfies a WIPI with $\LBI(s)$, then $\mathcal L$ satisfies a WPI with $\beta(s)=\LBI\left(s/2\right)$.}
\end{theorem}
\begin{proof}
    ection{
    For WIPI for $\mathcal {\hat L}$, with taking $t=0$, we have that 
    \begin{align*}
        \|f\|_{\ELL^2(\hat\mu)}^2 &\leq s\|\mathcal{\hat L} f\|_{\ELL^2(\hat\mu)}^2 + \LBI\left(s\right)\|f\|^2_\osc.
    \end{align*}
     For functions $P_t\tilde f\in \ELL^2(\mu)\cap \mathcal D(\mathcal L)$ and $t\geq 0$, we have $(P_t\tilde f)\circ \pi \in \ELL^2(\hat \mu)\cap \mathcal D(\hat{\mathcal L})$, where $\pi$ is defined in \Cref{Sec: Lift}. Then, we have
    \begin{align*}
        &\|(P_t\tilde f)\circ \pi\|_{\ELL^2(\hat\mu)}^2 \leq s\|\mathcal{\hat L} (P_t\tilde f)\circ \pi\|_{\ELL^2(\hat\mu)}^2 + \LBI\left(s\right)\|(P_t\tilde f)\circ \pi\|^2_\osc,\\
        \Longrightarrow
        &\|P_t\tilde f\|_{\ELL^2(\mu)}^2 \leq 2\cdot s\langle P_t\tilde f,  -\mathcal{L}P_t\tilde f\rangle_{\ELL^2(\mu)}+ \LBI\left(s\right)\|P_t\tilde f\|^2_\osc
    \end{align*}
    The second equality is due to the second-order lifting condition, i.e. $\langle\hat{\mathcal{L}}(P_t\tilde f)\circ \pi,\hat{\mathcal{L}} (P_t\tilde f)\circ \pi\rangle_{\ELL^2(\hat \mu)}=2\langle (P_t\tilde f) ,  -\mathcal{L}(P_t\tilde f)\rangle_{\ELL^2(\mu)}$.
    }
    
    ection{
    Let $s'=2s$, we have
    \begin{align*}
        \|P_t\tilde f\|_{\ELL^2(\mu)}^2 \leq  s'\langle P_t\tilde f,  -\mathcal{L}P_t\tilde f\rangle_{\ELL^2(\mu)}+ \beta(s')\|f\|^2_\osc
    \end{align*}
    with $\beta(s')=\LBI\left(s'/2\right)$.
    }
\end{proof}

\begin{proof}[Proof of \Cref{thm: main}]
\label{prf: thm: main}
    As $\hat{\mathcal{L}}$ satisfies a WFPI, it must also satisfy some WIFPI by \Cref{thm:order}. And it must also satisfy some WIPI by \Cref{thm:lift}.
    
    Hence, for $\hat{\mathcal L}$ and $f \in \ELL^2_0(\hat \mu)\cap \mathcal D(\hat{\mathcal L})$, we have
\begin{itemize}
    \item WFPI: $\LKF\left(\frac{    \int_0^T\left\| \hat P_t f\right\|_{\ELL^2(\hat\mu)}^2 \dif t}{\|f\|^2_\osc}\right)\leq \frac{\int_0^T\left\langle \hat P_t f,-\hat{\mathcal{L}}\hat P_t f\right\rangle_{\ELL^2(\hat\mu)} \dif t}{\|f\|^2_\osc}.$
    \item WIFPI: $\int_0^T\left\| \hat P_t f\right\|_{\ELL^2(\hat\mu)}^2 \dif t \leq s \cdot \int_0^T \|\hat{\mathcal{L}}\hat P_t f\|^2_{\ELL^2(\hat\mu)} \dif t+ \LBIF(s)\|f\|^2_\osc.$
    \item WIPI: $ \|\hat P_tf\|_{\ELL^2(\hat\mu)}^2 \leq s \cdot \|\mathcal{\hat L} \hat P_tf\|_{\ELL^2(\hat\mu)}^2 + \LBI(s)\|f\|^2_\osc.$
\end{itemize}

For $\mathcal{L}$ and $\tilde f \in \ELL^2_0(\mu)\cap \mathcal D(\mathcal L)$, we have
\begin{itemize}
    \item WPI: $\left\| P_t \tilde f \right\|_{\ELL^2(\mu)}^2 \leq s \cdot \left\langle P_t \tilde f,-{\mathcal{L}}P_t \tilde f\right\rangle_{\ELL^2(\mu)} + \beta^\star(s)\|f\|^2_\osc.$
\end{itemize}
    
    We denote the optimal $\beta$ for WFPI of $\mathcal L$ as $\BF^\star$. Hence \Cref{thm:reversible} gives 
    \begin{align*}
        \beta^\star(s) \leq \LBI(s/2)\quad \Longrightarrow \quad
        \OLBI(s) \geq \beta^\star(2s),
    \end{align*}
    where it is straightforward that $\beta^\star(2s) \in \mathcal{B}$.
    
    Next, by \Cref{thm:lift}, we have
    \begin{align*}
        \OLBI(s)\leq \frac{2}{T}\LBIF\left(\frac{s'-2T^2}{2}\right) \quad \Longrightarrow \quad
        \OLBIF(s) \geq \frac{T}{2}\OLBI(2T^2+2s)\geq \frac{T}{2}\beta^\star(4T^2+4s).
    \end{align*}
    where it is straightforward that $\frac{T}{2}\beta^\star(4T^2+4s) \in \mathcal{B}$.

    Then, under \Cref{asp: T}, \Cref{cor:order} shows $\OLKF \in \mathcal{C}$ and $\mathcal{\hat L}$ satisfies a WFPI with $\LKF\leq \OLKF$, where
    \begin{align*}
        \frac{(\OLKF(v))^2}{v}\leq \OLKIF(v) \quad \Longrightarrow \quad
        \OLKF(v) \leq \sqrt{\OLKIF(v) \cdot v},
    \end{align*}
    where $\OLKIF$ coresponds to $\OLBIF(s)$, and $\OLKIF \in \mathcal{C}$.
\end{proof}

\section{Examples}
\label{Sec: Examples}
\subsection{Exponential case}
\begin{theorem}
    Suppose that $(P_t)_{t\geq0}$ admits a spectral gap $\gap(\mathcal L)>0$, and that $(\hat P_t)_{t\geq0}$ is $T$-delayed exponentially contractive with rate $\nu$, then its decay rate $\nu$ satisfies
    \begin{align*}
        \frac{1}{\nu} \geq \sqrt{\frac{1}{4 \gap(\mathcal L)}-T^2}.
    \end{align*}
\end{theorem}
\begin{proof}
    Based on \Cref{rmk:strongtoweak}, $(P_t)_{t\geq0}$ satisfies WPI with $\beta(s)=\mathbbm{1}_{\{s < \gap(\mathcal L)^{-1}\}}$. 
    Therefore, we could choose $\beta^\star(s)\geq (1-\gap(\mathcal L)s)_+/4$. So $\beta^\star(s)=0$ when $s\geq \gap(\mathcal L)^{-1}$.

    \Cref{thm:lift} and \Cref{thm:reversible} show $(\hat P_t)_{t\geq0}$ satisfies a WIFPI with $\OLBIF(s)$ with $\OLBIF(s)\geq \frac{T}{2}\beta^\star(4T^2+4s)$. Let $\phi(s)=\frac{T}{2}\beta^\star(4T^2+4s)$, we have 
    \begin{align*}
    \phi(s)&=0, \quad \text{when } s\geq \gap(\mathcal{L})^{-1}/4-T^2,\\
    \OLKIF(v) &\leq \frac{v}{\gap(\mathcal{L})^{-1}/4-T^2}, \quad \text{when } 0\le v \leq T/2.
    \end{align*}
    Then, let $\mathcal {C'}=\{K^*(v)=Cv: C>0, 0\leq v \leq 1/4 \}$. By \Cref{thm: main}, we have
    \begin{align*}
        \OLKF(v) \leq  \sqrt{v \cdot \OLKIF(v)}&\leq \sqrt {\frac{4}{\gap(\mathcal L)^{-1}-4T^2}} \cdot v.
    \end{align*}
    Hence, as shown in \Cref{eq:KF}, $\nu \leq \sqrt {\frac{4}{\gap(\mathcal L)^{-1}-4T^2}}$, i.e.
    \begin{align*}
        \frac{1}{\nu} \geq \sqrt{\frac{1}{4 \gap(\mathcal L)}-T^2}.
    \end{align*}
\end{proof}

\begin{remark}
    Comparing with $\frac{1}{\nu} \geq \frac{1}{\sqrt{2\gap(\mathcal L)}}-T$ in \cite{eberle2024non}, our bound shows the same square-root order of $\gap(\mathcal L)$.
\end{remark}

\subsection{Polynomial case}
\begin{theorem}
    Suppose that $(P_t)_{t\geq0}$ admits a WPI with $\beta^\star(s)\geq c_0s^{-c_1}  \wedge c_2$, where $c_0,c_1>0$ and $0<c_2<1/4$, and $(\hat P_t)_{t\geq0}$ satisfies a WFPI with $\LKF$ and $T\in(0,\infty)$. Let $\gamma^\star$ and $\widehat{\gamma}_F^\star$ be the respective convergence bound functions as defined above in \Cref{thm:WPI_convergence}. For $t>t_0>0$, we have
    \begin{align*}
        \gamma^\star (t)&\geq M t^{-c_1},\\
        \widehat{\gamma}_\mathrm F^\star(t)&\geq M' t^{-2c_1},
    \end{align*}
    where $M,M'$ are universal constants.
\end{theorem}

\begin{proof}
    Since $\beta^\star(s)\geq c_0s^{-c_1}\wedge c_2$, and $\OLBIF$ is given by WIFPI for $\mathcal{\hat L}$. \Cref{thm:lift} and \Cref{thm:reversible} show, 
\begin{align*}
    \frac{2}{T}\OLBIF\left(\frac{s'-2T^2}{2}\right)&\geq \beta^\star(s),\\
    \OLBIF(s)&\geq \frac{T}{2}c_0(4s+4T^2)^{-c_1}\wedge c_2T/2.
\end{align*}
Hence, there exists a uniform constant $C_0, C_1$, where $C_1<T/4$, such that $\OLBIF(s)\geq C_0(c_0s^{-c_1})\wedge C_1$. 
Then, we have $\OLKIF(v)\leq C (C_0, c_1) v^{1+c_1^{-1}}$, as $0 \leq v\leq T/4$.

Let $\mathcal{C'}=\{K^*(v): K^*(v)=Cv^{1+c_1^{-1}}, C>0,c_1>0, 0\leq v\leq T/4\}$.
By \Cref{thm: main}, when $0 \leq v\leq T/4$,
\begin{align*}
    \LKF(v) &\leq \sqrt{v \cdot \OLKIF(v)} \leq C'(C_0, c_1) v^{1+(2c_1)^{-1}}.
\end{align*}

It is easy to show there exist universal constants $M,M'>0$ and $t_0>0$ such that, for any $t>t_0$,  
\begin{align*}
    \gamma^\star (t)&\geq M t^{-c_1}
    \quad \text{and } \quad\widehat{\gamma}_\mathrm  F^\star(t)\geq M' t^{-2c_1}.
\end{align*}
\end{proof}

\subsection{Example: Cauchy distribution}
    We consider a 1-d Cauchy distribution as the target distribution. 
    \cite{eberle2024non} shows that underdamped Langevin and the Zig-Zag process are both second-order lifting of the overdamped Langevin diffusion.

    In total variation, \cite{brevsar2026subexponential} obtained the lower bound of order $Ct^{-1/2}$ for the convergence of the overdamped Langevin diffusion, and the same order holds for the underdamped Langevin diffusion, while Corollary 1 of \cite{vasdekis2022note} gives an upper bound of order $C't^{-1}$ for the Zig-Zag process. Thus, in this example, a speed-up associated with second-order lifting can be observed in total variation distance in the subgeometric setting. However, this does not imply that every second-order lifting achieves this speed-up.
    
    For $\ELL^2$ convergence, \cite{brigati2024explicit} provides upper bounds of the same order for the overdamped and underdamped Langevin diffusions. The situation for the Zig-Zag process is less clear: an $\ELL^2$ convergence rate upper bound of $\gamma(t)=t^{-1/8}$ as defined in \cref{eq:convergence rate}, is given in \cite{Andrieu2021}, which does not demonstrate any acceleration; however this is strongly believed not to be sharp. Therefore, there is no direct evidence at present that the acceleration predicted by our theoretical bound can be attained in $\ELL^2$ for the Zig-Zag process.

\begin{figure}[H]
    \centering

    \begin{subfigure}{0.48\textwidth}
        \centering
        \includegraphics[width=\textwidth]{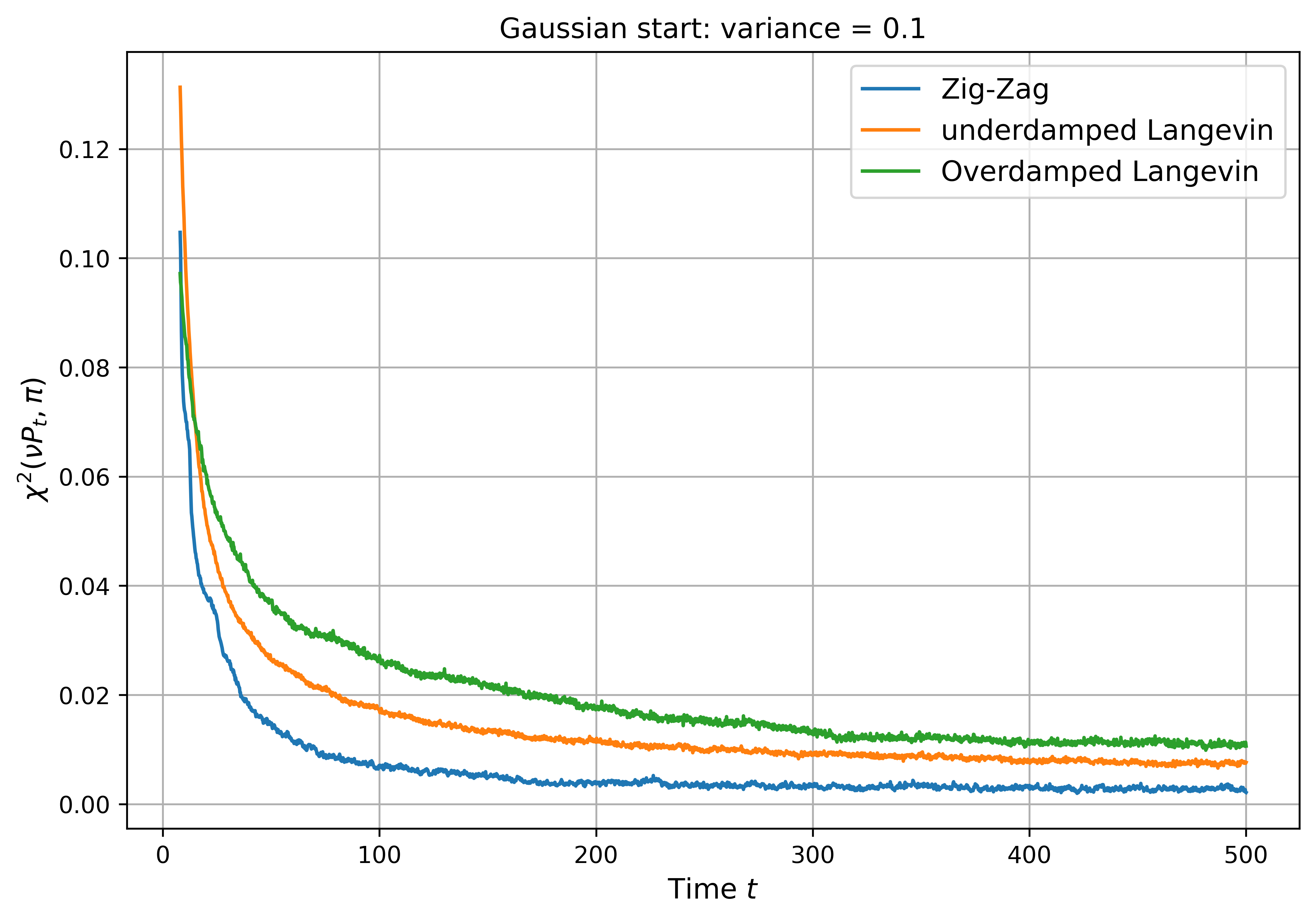}
        \caption{$\chi^2$-distance: The plots use a thinning factor of 20.}
        \label{fig:fig1}
    \end{subfigure}
    \hfill
    \begin{subfigure}{0.48\textwidth}
        \centering
        \includegraphics[width=\textwidth]{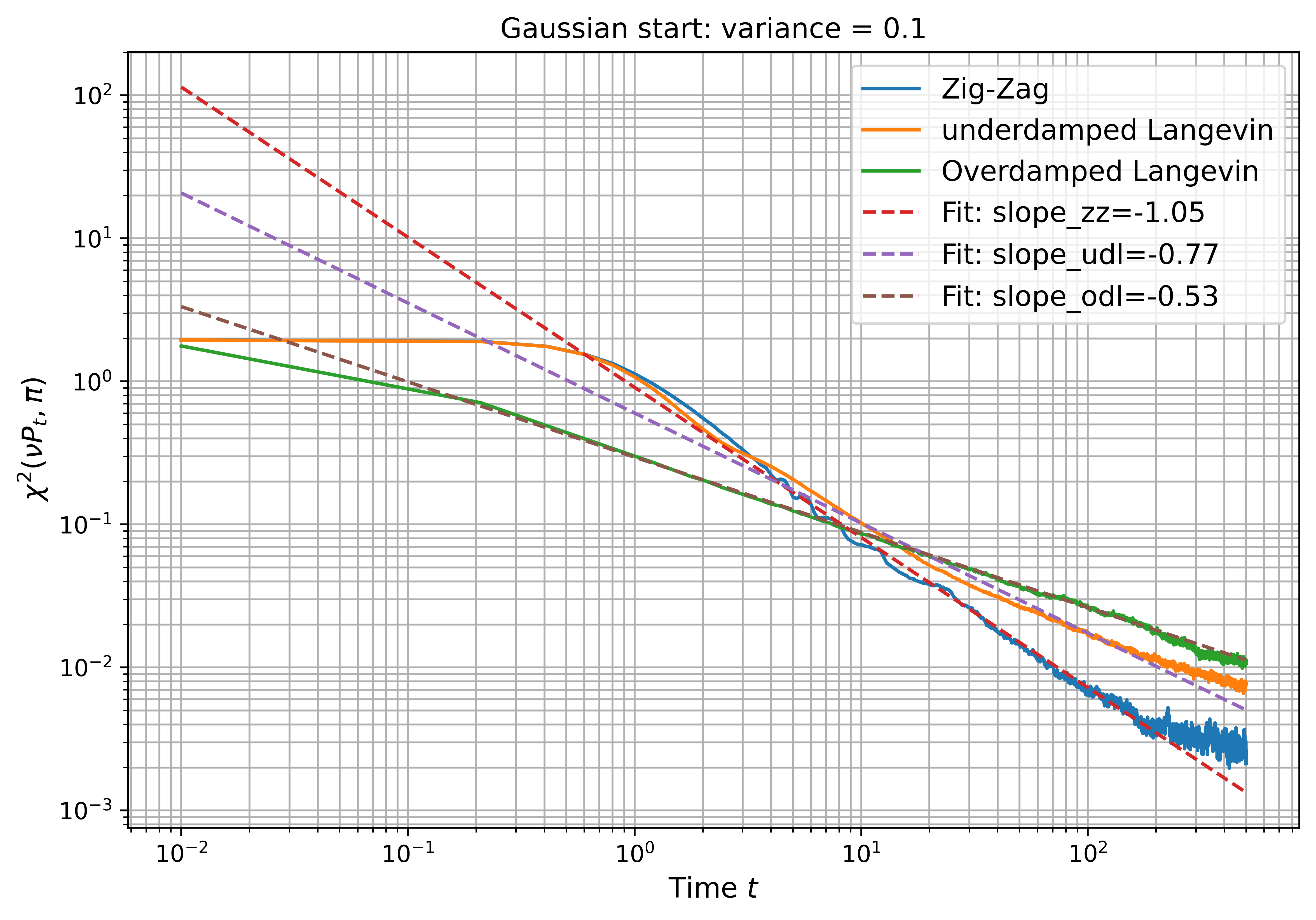}
        \caption{The corresponding log--log scale: The slopes are estimated by linear regression over iterations 2–1000 after thinning.}
        \label{fig:fig2}
    \end{subfigure}

    \vspace{0.5cm}

    \begin{subfigure}{0.48\textwidth}
        \centering
        \includegraphics[width=\textwidth]{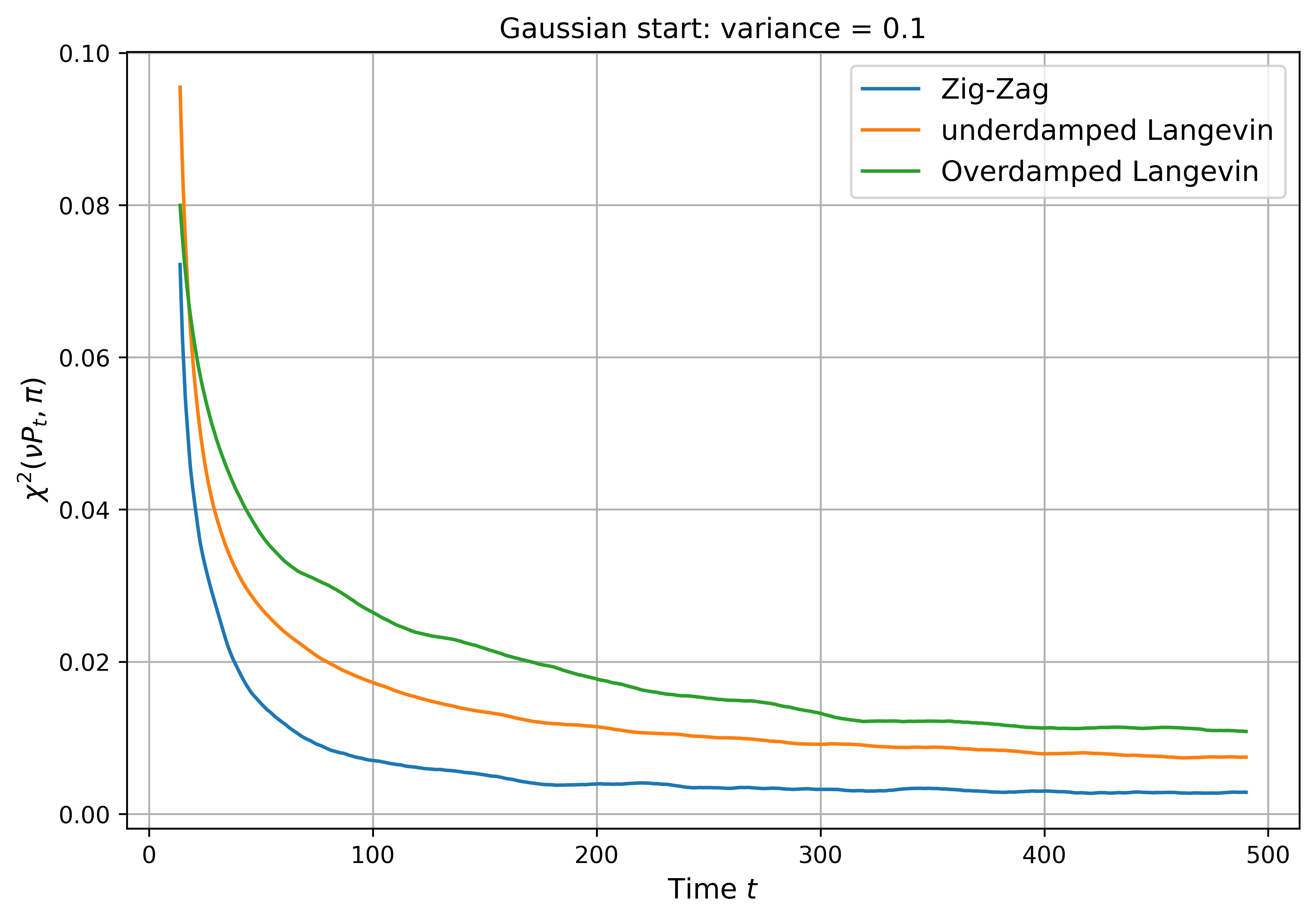}
        \caption{The corresponding figure with smoothing window length 100.}
        \label{fig:fig3}
    \end{subfigure}
    \hfill
    \begin{subfigure}{0.48\textwidth}
        \centering
        \includegraphics[width=\textwidth]{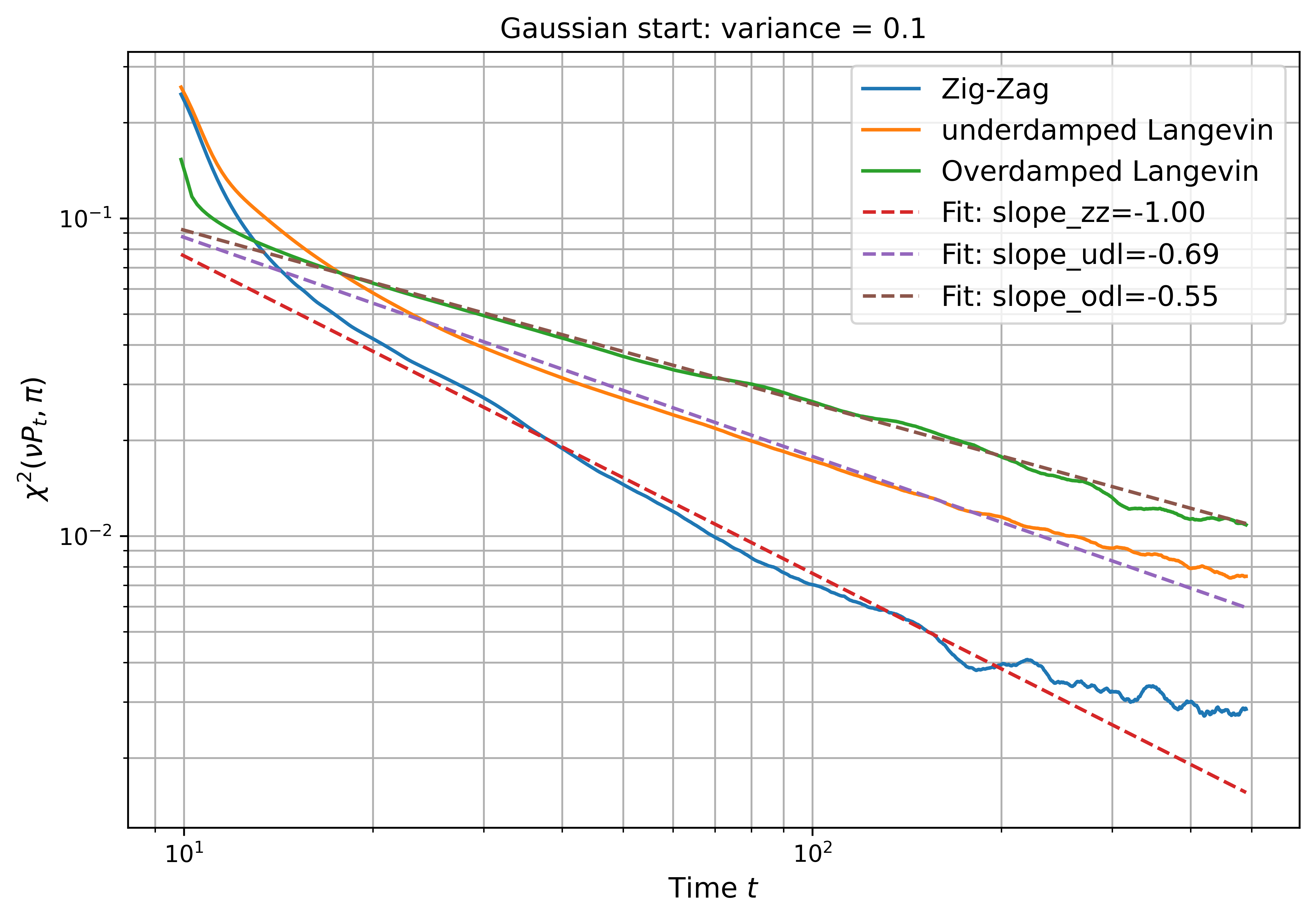}
        \caption{The corresponding smoothed log--log scale: The slopes are estimated by linear regression over smoothed iterations 10-1200 after thinning.}
        \label{fig:fig4}
    \end{subfigure}

    \caption{{Decay of the $\chi^2$-distance and its asymptotic decay rate}}
    \label{fig:four_figures}
\end{figure}

    We numerically compare Langevin dynamics and the one-dimensional Zig-Zag process {(without refreshment)} targeting a standard Cauchy distribution. Starting from a Gaussian distribution centered at 0 with variance 0.1, {we estimate the $\chi^2$ distances of the joint distribution to stationarity using empirical histograms with target-probability bins $80$, $80 \times 8 $ and $80 \times 2 $ for overdamped Langevin, underdamped Langevin and Zig-Zag, respectively in \Cref{fig:four_figures}. 
    To run the simulation we generated 40,000, $40,000\times 4$, $40,000\times 1.5$ particles for overdamped Langevin, underdamped Langevin and Zig-Zag, respectively, and ran the process for 50,000 iterations. We investigate the resulting polynomial decay rates. }

    From \Cref{fig:fig2} and \Cref{fig:fig4}, the fitted slopes indicate that an acceleration due to lifting can be observed for some initial distributions. Based on these numerical results, we conjecture that, for the one-dimensional Cauchy target considered here, the sharp $\ELL^2 $ convergence rates are $t^{-1/2}$ for the overdamped Langevin diffusion and $t^{-1}$ for the Zig-Zag process.

\section*{AI Statement}
All mathematical results -- the formulations, original proofs, and scientific contributions -- are the original work of the authors. Generative AI tools were only used in final stages of manuscript preparation for technical checks and language editing.

\section{Appendix}
\begin{proof}[Proof of \Cref{thm:WPI_convergence}]
    \label{prf:thm:WPI_convergence}
    Let $f \in \ELL^2_0(\mu)$ and $f=P_0 f$. Then we have
    \begin{align*}
     \frac{\dif}{\dif t}\left\|P_t f\right\|_{\ELL^2(\mu)}^2 =2 \left\langle P_t f, \mathcal{L} P_t f\right\rangle_{\ELL^2(\mu)}.
    \end{align*}
    Now, applying the optimised weak Poincaré inequality,
    \begin{align*}
        K^*\left(\frac{\left\|P_t f\right\|_{\ELL^2(\mu)}^2}{\|f\|^2_\osc}\right) &\leq \frac{\left\langle P_t f,-{\mathcal{L}} P_t f\right\rangle_{\ELL^2(\mu)}}{\|f\|^2_\osc}= -\frac{1}{2}\frac{\dif}{\dif t}\frac{\left\|P_t f\right\|_{\ELL^2(\mu)}^2}{\|f\|^2_\osc}\\
        \Rightarrow\frac{\dif}{\dif t}\frac{\left\|P_t f\right\|_{\ELL^2(\mu)}^2}{\|f\|^2_\osc}&\leq -2K^*\left(\frac{\left\|P_t f\right\|_{\ELL^2(\mu)}^2}{\|f\|^2_\osc}\right).
    \end{align*}
    Let $F(x)=\int_x^{1/4}\frac{\dif v}{2K^*(v)}=-\int_{1/4}^x\frac{\dif v}{2K^*(v)}$, and $g(t)=\frac{\left\|P_t f\right\|_{\ELL^2(\mu)}^2}{\|f\|^2_\osc}$. A Bihari–Lasalle argument yields
    \begin{align*}
        F(g(t))-F(g(0)) &\geq \int_0^t \dif s,\\
        \Rightarrow\frac{\left\|P_t f\right\|_{\ELL^2(\mu)}^2}{\|f\|^2_\osc} &\leq F^{-1}\left(t+F\left(\frac{\left\|f\right\|_{\ELL^2(\mu)}^2}{\|f\|^2_\osc}\right)\right),
    \end{align*}
    since $F^{-1}$ is decreasing. Then we have
    \begin{align*}
        \|P_t f\|_{\ELL^2(\mu)}^2\leq F^{-1}\left(t+F\left(\frac{\left\|f\right\|_{\ELL^2(\mu)}^2}{\|f\|^2_\osc}\right)\right)\cdot \|f\|^2_\osc \leq F^{-1}(t)\cdot \|f\|^2_\osc.
    \end{align*}
\end{proof}

\begin{proof}[Proof of \Cref{thm:WFPI_convergence}]
\label{prf:thm:WFPI_convergence}

Let $f \in \ELL^2_0(\mu)$. Following \cite{eberle2024non}, we have
    \begin{align*}
& \frac{\dif}{\dif t} \int_t^{t+T}\left\|P_s f\right\|_{\ELL^2(\mu)}^2 \dif s=\left\|P_{t+T} f\right\|_{\ELL^2(\mu)}^2-\left\|P_t f\right\|_{\ELL^2(\mu)}^2 \\
& \quad=\int_t^{t+T} \frac{\dif}{\dif s}\left\|P_s f\right\|_{\ELL^2(\mu)}^2 \dif s=2 \int_t^{t+T}\left\langle P_s f, \mathcal{L} P_s f\right\rangle_{\ELL^2(\mu)} \dif s.
    \end{align*}
    By the semigroup property, the WFPI implies
    \begin{align*}
        K^*\left(\frac{    \int_0^T\left\| P_{t+s}f\right\|_{\ELL^2(\mu)}^2 \dif t}{\|f\|^2_\osc}\right)&\leq \frac{\int_0^T\left\langle P_{t+s}f,-{\mathcal{L}} P_{t+s}f\right\rangle_{\ELL^2(\mu)} \dif t}{\|f\|^2_\osc},
    \end{align*}
    so that 
    \begin{align}
    \label{eq1}
        \frac{\frac{\dif}{\dif t}\int_t^{t+T}\|P_s f\|^2\dif s}{\|f\|^2_\osc}\leq - 2 K^*\left(\frac{    \int_t^{T+t}\left\| P_{s}f\right\|_{\ELL^2(\mu)}^2 \dif s}{\|f\|^2_\osc}\right).
    \end{align}
    Let $F(x)=\int_x^{T/4}\frac{\dif v}{2K^*(v)}=-\int_{T/4}^x\frac{\dif v}{2K^*(v)}$, and $g(t)=\frac{    \int_t^{T+t}\left\| P_{s}f\right\|_{\ELL^2(\mu)}^2 \dif s}{\|f\|^2_\osc}$.
    
    A Bihari–Lasalle argument yields
    \begin{align*}
        F_a(g(t))-F_a(g(0)) &\geq \int_0^t \dif s,\\
        \frac{    \int_t^{T+t}\left\| P_{s}f\right\|_{\ELL^2(\mu)}^2 \dif s}{\|f\|^2_\osc} &\leq F^{-1}\left(t+F\left(\frac{    \int_0^{T}\left\| P_{s}f\right\|_{\ELL^2(\mu)}^2 \dif s}{\|f\|^2_\osc}\right)\right),
    \end{align*}
    since $F^{-1}$ is decreasing.

    Then for $t\geq T$,
    \begin{align*}
         \frac{\left\| P_{t}f\right\|_{\ELL^2(\mu)}^2 }{\|f\|^2_\osc}&\leq \frac{1}{T} \frac{\int_{t-T}^t\left\| P_{s}f\right\|_{\ELL^2(\mu)}^2 \dif s}{\|f\|^2_\osc}\\
         &\leq \frac{1}{T} F^{-1}\left(t-T+F\left(\frac{    \int_0^{T}\left\| P_{s}f\right\|_{\ELL^2(\mu)}^2 \dif s}{\|f\|^2_\osc}\right)\right)\\
         &\leq \frac{1}{T} F^{-1}\left(t-T\right).
    \end{align*}
\end{proof}

    \begin{proof}[Proof of \Cref{thm: from rate to WFPI}]
\label{prf: thm: from rate to WFPI}
 Let $g(t)=\frac{    \int_t^{T+t}\left\| P_{s}f\right\|_{\ELL^2(\mu)}^2 \dif s}{\|f\|^2_\osc}$. As $F$ is decreasing, continuous, and divergent at $0$, we have 
 \begin{align*}
    \frac{F(g(t))-F(g(0))}{t}\geq 1 \quad 
    \Rightarrow  \quad\lim_{t \rightarrow 0^+}\frac{F(g(t))-F(g(0))}{t}\geq 1.
 \end{align*}
 This yields $(\mathrm{D}F)(g(0))g'(0)\geq 1$. Since $\frac{\dif}{\dif t} \int_t^{t+T}\left\|P_s f\right\|_{\ELL^2(\mu)}^2 \dif s=2 \int_t^{t+T}\left\langle P_s f, \mathcal{L} P_s f\right\rangle_{\ELL^2(\mu)} \dif s$ and $(\mathrm{D}F)(g(0))< 0$,
 \begin{align*}
     \frac{ \int_0^{T}\left\langle P_s f, \mathcal{-L} P_s f\right\rangle_{\ELL^2(\mu)} \dif s}{\|f\|^2_\osc} \geq -\frac{1}{2 }(\mathrm{D}F)^{-1}\left(\frac{\int_0^{T}\left\|P_s f\right\|_{\ELL^2(\mu)}^2 \dif s}{\|f\|^2_\osc}\right).
 \end{align*}

 As such, it suffices to prove that the mapping $\KF(v):= -\frac{1}{2 }(\mathrm{D}F)^{-1}(v)$ is $0$ at $0$, increasing, and convex.
\end{proof}

\begin{proof}[Proof of \Cref{thm:C}]
\label{prf:thm:C}
For any $K^*\in \mathcal C_0$, let $K(u):=\sup_{0\leq v < a} \left\{u\cdot v-K^*(v)\right\}$, for $u \geq 0$, and let $\beta(s):=s\cdot K(s^{-1})$ for $s>0$. As shown in the proof of Proposition 4(e) in \cite{andrieu2026weak}, $\beta(s)\rightarrow 0$ as $s\rightarrow \infty$, and $\beta(s)$ is decreasing. Then $K(u)=u\beta(1/u)$, $K^*(v)=\sup_{0\leq u}\left\{u\cdot v-K(v)\right\}$ as $v\in [0,a)$, since $K^*$ is convex and strictly increasing. 
\end{proof}

%% file: ref.bib
@article{eberle2024non,
  title={{Non-reversible lifts of reversible diffusion processes and relaxation times}},
  author={Eberle, Andreas and L{\"o}rler, Francis},
  journal={Probability Theory and Related Fields},
  year={2026},
  volume={194},
  pages={173-203}
}

@article{Andrieu2021,
author = {Andrieu, Christophe and Dobson, Paul and Wang, Andi Q.},
doi = {10.1214/21-EJP643},
issn = {1083-6489},
journal = {Electronic Journal of Probability},
number = {none},
pages = {1--26},
publisher = {Institute of Mathematical Statistics and Bernoulli Society},
title = {{Subgeometric hypocoercivity for piecewise-deterministic Markov process Monte Carlo methods}},
volume = {26},
year = {2021}
}

@article{vasdekis2022note,
  title={{A note on the polynomial ergodicity of the one-dimensional Zig-Zag process}},
  author={Vasdekis, Giorgos and Roberts, Gareth O},
  journal={Journal of Applied Probability},
  volume={59},
  number={3},
  pages={895--903},
  year={2022},
  publisher={Cambridge University Press}
}

@article{andrieu2022comparison,
  title={{Comparison of Markov chains via weak Poincar{\'e} inequalities with application to pseudo-marginal MCMC}},
  author={Andrieu, Christophe and Lee, Anthony and Power, Sam and Wang, Andi Q},
  journal={The Annals of Statistics},
  volume={50},
  number={6},
  pages={3592--3618},
  year={2022},
  publisher={Institute of Mathematical Statistics}
}

@article{andrieu2026weak,
  title={{Weak Poincar{\'e} Inequalities for Markov chains: theory and applications}},
  author={Andrieu, Christophe and Lee, Anthony and Power, Sam and Wang, Andi Q},
  journal={The Annals of Applied Probability},
  volume={36},
  number={1},
  pages={46--107},
  year={2026},
  publisher={Institute of Mathematical Statistics}
}

@article{brevsar2026subexponential,
  title={{Subexponential lower bounds for f-ergodic Markov processes}},
  author={Bre{\v{s}}ar, Miha and Mijatovi{\'c}, Aleksandar},
  journal={Probability Theory and Related Fields},
  volume={194},
  number={1},
  pages={997--1054},
  year={2026},
  publisher={Springer}
}

@article{brigati2024explicit,
  title={{Explicit convergence rates of underdamped Langevin dynamics under weighted and weak Poincar\'e--Lions inequalities}},
  author={Brigati, Giovanni and Stoltz, Gabriel and Wang, Andi Q and Wang, Lihan},
  journal={The Annals of Applied Probability},
  year={2026}
}

@article{rockner2001weak,
  title={{Weak Poincar{\'e} inequalities and L2-convergence rates of Markov semigroups}},
  author={R{\"o}ckner, Michael and Wang, Feng-Yu},
  journal={Journal of Functional Analysis},
  volume={185},
  number={2},
  pages={564--603},
  year={2001},
  publisher={Elsevier}
}

@article{huang2026bernstein,
  title={{Bernstein-type inequalities for Markov chains and Markov processes: A simple and robust proof}},
  author={Huang, De and Li, Xiangyuan},
  journal={Bernoulli},
  volume={32},
  number={2},
  pages={1267--1284},
  year={2026},
  publisher={Bernoulli Society for Mathematical Statistics and Probability}
}

@article{lelievre2013optimal,
  title={{Optimal non-reversible linear drift for the convergence to equilibrium of a diffusion}},
  author={Lelievre, Tony and Nier, Francis and Pavliotis, Grigorios A},
  journal={Journal of Statistical Physics},
  volume={152},
  number={2},
  pages={237--274},
  year={2013},
  publisher={Springer}
}

@article{bouchard2018bouncy,
  title={{The bouncy particle sampler: A nonreversible rejection-free Markov chain Monte Carlo method}},
  author={Bouchard-C{\^o}t{\'e}, Alexandre and Vollmer, Sebastian J and Doucet, Arnaud},
  journal={Journal of the American Statistical Association},
  volume={113},
  number={522},
  pages={855--867},
  year={2018},
  publisher={Taylor \& Francis}
}

@article{eberle2025convergence,
  title={{Convergence of non-reversible Markov processes via lifting and flow Poincar{\'e} inequality}},
  author={Eberle, Andreas and Guillin, Arnaud and Hahn, Leo and L{\"o}rler, Francis and Michel, Manon},
  journal={arXiv preprint arXiv:2503.04238},
  year={2025}
}

@article{bierkens2019zig,
  title={{The Zig-Zag process and super-efficient sampling for Bayesian analysis of big data}},
  author={Bierkens, Joris and Fearnhead, Paul and Roberts, Gareth},
  journal={The Annals of Statistics},
  volume={47},
  number={3},
  pages={1288--1320},
  year={2019}
}

@inproceedings{chen1999lifting,
  title={{Lifting Markov chains to speed up mixing}},
  author={Chen, Fang and Lov{\'a}sz, L{\'a}szl{\'o} and Pak, Igor},
  booktitle={Proceedings of the thirty-first annual ACM symposium on Theory of computing},
  pages={275--281},
  year={1999}
}

@article{diaconis2000analysis,
  title={{Analysis of a nonreversible Markov chain sampler}},
  author={Diaconis, Persi and Holmes, Susan and Neal, Radford M},
  journal={Annals of Applied Probability},
  pages={726--752},
  year={2000},
  publisher={JSTOR}
}

@article{xu2026singular,
  title={{Singular-value gap of nonreversible Markov processes}},
  author={Xu, Ruochuan},
  journal={arXiv preprint arXiv:2606.01683},
  year={2026}
}

@article{eberle2026relaxation,
  title={{Relaxation times of non-reversible Markov processes}},
  author={Eberle, Andreas and L{\"o}rler, Francis},
  journal={arXiv preprint arXiv:2607.10801},
  year={2026}
}

@article{chatterjee2025spectral,
  title={{Spectral gap of nonreversible Markov chains}},
  author={Chatterjee, Sourav},
  journal={The Annals of Applied Probability},
  volume={35},
  number={4},
  pages={2644--2677},
  year={2025},
  publisher={Institute of Mathematical Statistics}
}

@article{dolbeault2009hypocoercivity,
  title={Hypocoercivity for kinetic equations with linear relaxation terms},
  author={Dolbeault, Jean and Mouhot, Cl{\'e}ment and Schmeiser, Christian},
  journal={Comptes Rendus. Math{\'e}matique},
  volume={347},
  number={9-10},
  pages={511--516},
  year={2009}
}

@article{villani2006hypocoercivity,
  title={Hypocoercivity},
  author={C\'edric Villani},
  journal={arXiv preprint math/0609050},
  year={2006}
}
